\documentclass[]{interact}

\usepackage{hyperref}  
\usepackage{subcaption}
\usepackage{float}

\begin{document}


\title{Bayesian Quantile Regression for Longitudinal Count Data}

\author{
\name{Sanket Jantre\textsuperscript{a}\thanks{Contact: jantresa@msu.edu}}
\affil{\textsuperscript{a}Department of Statistics and Probability, Michigan State University, MI, USA}
}

\maketitle

\begin{abstract}
This work introduces Bayesian quantile regression modeling framework for the analysis of longitudinal count data. In this model, the response variable is not continuous and hence an artificial smoothing of counts is incorporated. The Bayesian implementation utilizes the normal-exponential mixture representation of the asymmetric Laplace distribution for the response variable. An efficient Gibbs sampling algorithm is derived for fitting the model to the data. The model is illustrated through simulation studies and implemented in an application drawn from neurology. Model comparison demonstrates the practical utility of the proposed model. 
\end{abstract}


\section{Introduction}
\label{Section1}
Research on count data models has accelerated following the work on multiple regression analysis of a Poisson process by \cite{Jorgenson-1961}. The focus of successive literature lies on semiparametric estimation of conditional mean of count variables underneath pseudolikelihood framework. However, when the underlying response variable distribution is asymmetric, mean estimates become inadequate for inferences on the parameters of interest. This shortcoming prompted the research on fully parametric models describing the complete conditional distribution of parameters allowing to investigate the impact of the covariates on every aspect of the conditional distribution. Nonetheless, the strong parametric assumption and non-robust nature make these techniques unappealing.

More recent approach to the modeling and analyzing count data has involved the use of quantile regression techniques introduced by \cite{Koenker-Basset-1978}. The central theme of their work is to estimate the conditional quantile functions. The idea of median regression involving minimization of sum of absolute residuals was introduced by Boscovich in the mid-18th century. The robustness of median regression over mean regression in the presence of even small number of outlying observations is discussed in literature at length, remarkably by Kolmogorov \cite{Shiryayev-1992}. The commonplace assumption of Gaussian errors is often dubious and might result in surprising estimates of the coefficients of explanatory variables. Conversely, the quantile regression methods are able to handle the heteroscedasticity in the errors as well. A detailed guide on quantile regression and its properties is furnished by \cite{Koenker-2005}.

Various extensions to the quantile regression methods have been proposed in classical and Bayesian literature. Most of the developed techniques are primarily focused on continuous response data. The classical strand includes the simplex algorithm \cite{Dantzig-1963,Dantzig-Thapa-1997,Dantzig-Thapa-2003,Barrodale-roberts-1973,Koenker-dOrey-1987}, the interior point algorithm \cite{Karmarkar-1984,Mehrotra-1992,Portnoy-Koenker-1997}, the smoothing algorithm \cite{Madsen-Nielsen-1993,Chen-2007} and metaheuristic algorithms \cite{Rahman-2013}. The crucial departure from continuous data setting in classical models was the work of \cite{Manski-1975,Manski-1985} for binary and multinomial models using median regression. Later, \cite{Powell-1984,Powell-1986} analyzed censored data using quantile regression, \cite{Lee-1992} studied median regression for discrete ordered responses and \cite{Machado-Silva-2005} gave quantile regression model for count data. 

On the other hand, Bayesian quantile regression pivoted on Markov chain Monte Carlo (MCMC) methods employing asymmetric Laplace (AL) likelihoods was introduced by \cite{Yu-Moyeed-2001}. \cite{Kozumi-Kobayashi-2011} proposed the normal-exponential mixture representation of the AL distribution and developed a Gibbs sampling algorithm. The AL likelihood has been used with or without the normal-exponential mixture to develop Bayesian estimation for quantile regression in Tobit models \cite{Yu-Stander-2007,Kozumi-Kobayashi-2011}, censored models \cite{Reich-Smith-2013}, binary models \cite{Hewson-Yu-2008,Benoit-Poel-2010}, count data models \cite{Lee-Neocleous-2010}, ordinal models \cite{Rahman-2016} and longitudinal data models \cite{Geraci-Bottai-2007,Luo-et-al-2012}. More recently, few works have emerged which propose Bayesian quantile regression analysis for discrete outcomes \cite{Rahman-Karnawat-2019,Alhamzawi-Ali-2020,Bresson-et-al-2021}.

Moving on to the longitudinal data models, \cite{Koenker-2004} presented $l_1$ penalization approach for the classical random effects estimator. In his model, an appropriate choice of degree of shrinkage or regularization term is required in order to control variability induced by random effects. Subsequently, likelihood based approach for quantile regression in longitudinal data models was proposed by \cite{Geraci-Bottai-2007} which makes use of AL density for likelihood to automate the choice of penalization level. They employed a Monte Carlo expectation maximization (EM) algorithm to estimate the model parameters. \cite{Luo-et-al-2012} made use of the normal-exponential mixture representation of AL density, presented in \cite{Kozumi-Kobayashi-2011}, to develop a random walk Metropolis Hastings (MH) algorithm and a Gibbs sampling algorithm for continuous longitudinal data. Furthermore, \cite{Rahman-2019} and \cite{Alhamzawi-2018} used the similar framework to develop efficient Gibbs samplers for binary and ordinal longitudinal data respectively. Similarly, \cite{Ghasemzadeh-et-al-2018, Ghasemzadeh-et-al-2020} have fitted a Bayesian quantile regression joint model to ordinal and continuous longitudinal data and used Gibbs sampling for posterior inference. In this work, we propose a Bayesian quantile regression for longitudinal count data models employing the normal-exponential mixture for AL likelihood.

Longitudinal data models commonly arise in economics, finance, epidemiology, clinical trials and sociology. Some of these data may involve intensity of event recurrence at a given time as responses for each subject. Poisson distribution properly models such response variables most of the times. Aforesaid data is referred in literature as panel count data and predominantly observed in demographic studies, clinical trials and industrial reliability \cite{Kalbfleisch-Lawless-1985,Gaver-OMuircheataigh-1987,Thall-Lachin-1988,Thall-1988,Sun-Kalbfleisch-1995}. Effects of specific regressors on the underlying counting process are of practical importance in these regression analysis studies. Bayesian quantile regression approach to panel data is an active area of research and with this work we attempt to further the analysis of panel count data using Bayesian techniques.

Panel data involves repeatedly measured responses for independent subjects over time, hence serial correlation is naturally observed amongst the measures from the same subject. In such scenario, mixed models with random effects were popularized by \cite{Laird-Ware-1982}. Introduction of random effects to account for the within-subject variability avoids bias in parameter estimates, takes care of overdispersion in responses, and can assist in improving the model fit as well. 

The rest of this paper is organized as follows. Section \ref{Section2} introduces quantile regression and establishes relationship between quantile regression and asymmetric Laplace distribution. In Section \ref{Section3}, we discuss longitudinal count data models and their composition in terms of Bayesian quantile regression. This section also proposes a hierarchical Bayesian quantile regression model for longitudinal count data and introduces an efficient Gibbs sampling algorithm. Section \ref{Section4} presents two Monte Carlo simulation studies and a real world application drawn from neurology. A brief discussion and conclusion is provided in Section \ref{Section5}. Technical derivations of conditional posterior densities are presented in the appendix.


\section{Quantile Regression and Asymmetric Laplace Distribution} 

\label{Section2} 


If we take $Y$ and $X$ random variables and $p\in (0,1)$ be the quantile level of the conditional distribution of $Y$, then a linear conditional quantile function is denoted as follows
\begin{equation} \label{quant_fun}
Q_{y_i}(p |x_i) \equiv F^{-1}(p) = x_i^T {\beta}_p, \hspace{1cm} i = 1,...,n, 
\end{equation} where $Q(.)$ is the inverse cdf of $y_i$ conditional on $x_i$ and ${\beta}_p \in \mathbb{R}^{k}$ is a vector of unknown fixed parameters of length $k$.

\begin{figure}[h]
\includegraphics[scale=0.5]{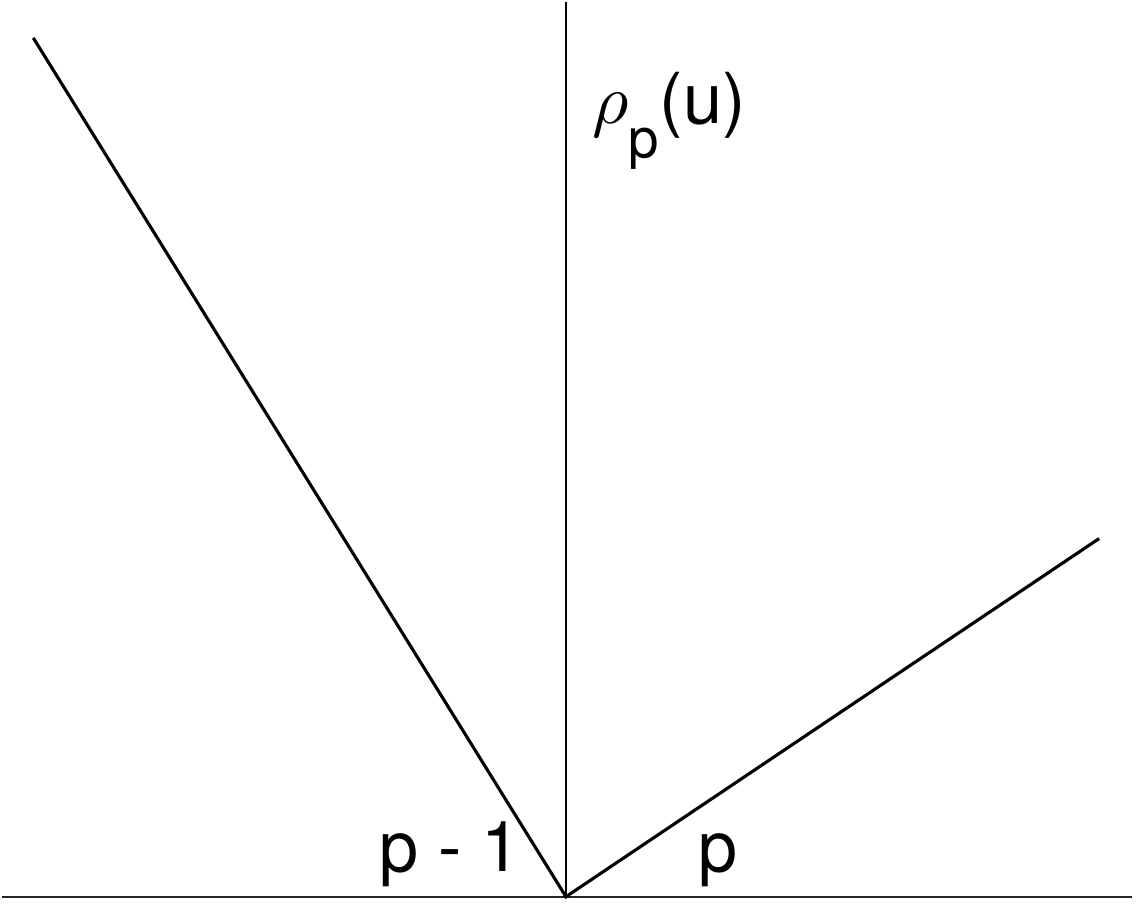}
\centering
\caption{Quantile regression check or loss function}
\label{quant_check_fun}
\end{figure}

We consider the quantile regression approach for a linear model given by
\begin{equation} \label{quant_lin_model}
y_i = x_i^T {\beta}_p + {\epsilon}_i, \hspace{1cm} i = 1,...,n,
\end{equation}
where ${\epsilon}_i$ is the error term with pdf $f(.)$ and is restricted to have its $p^\text{th}$ quantile to be zero.
\begin{equation*}
\int _{-\infty} ^ 0 f({\epsilon}_i) d{\epsilon}_i = p. 
\end{equation*}
Classical literature often does not mention the error density $f(.)$ and estimation through quantile regression proceeds by minimizing the following objective function
\begin{equation} \label{obj_fun}
\min_{{\beta}_p \in \mathbb{R}^k} \sum_{i=1}^n \rho_p (y_i - x_i^T \beta_p) 
\end{equation}
where $\rho_p(.)$ is the check function or quantile loss function defined by
\begin{equation*}
\rho_p(u) = u . \{p - I(u<0)\}
\end{equation*}
here $I(.)$ is an indicator function. This check function is not differentiable at zero, see Figure~\ref{quant_check_fun}. Classical literature employs linear programming techniques such as the simplex algorithm, the interior point algorithm, the smoothing algorithm or metaheuristic algorithms to obtain quantile regression estimates for ${\beta}_p$. The statistical programming language R makes use of \texttt{quantreg} package \cite{Koenker-2017} to implement quantile regression techniques whilst confidence intervals are obtained via bootstrapping methods \cite{Koenker-2005}.

Median regression in Bayesian setting has been considered by \cite{Walker-Mallick-1999} and \cite{Kottas-Gelfand-2001}. In quantile regression, a link between maximum-likelihood theory and minimization of the sum of check functions, given in \autoref{obj_fun}, is provided by asymmetric Laplace distribution (ALD) \cite{Koenker-Machado-1999,Yu-Moyeed-2001}. This distribution has location parameter $\mu$, scale parameter $\sigma$ and skewness parameter $p$. Further details regarding the properties of this distribution are specified in \cite{Yu-Zhang-2005}. If $Y {\sim}$ ALD$(\mu,\sigma,p)$, then its probability distribution function is given by
\begin{equation*}
f(y|\mu,\sigma,p) = \frac{p(1-p)}{\sigma} \exp \left \{- \rho_p \left (\frac{y-\mu}{\sigma}\right ) \right \} 
\end{equation*}
As discussed in \cite{Yu-Moyeed-2001}, using the above skewed distribution for errors provides a way to cope the problem of Bayesian quantile regression effectively. According to them, any reasonable choice of prior, even an improper prior, generates a posterior distribution for $\beta_p$. Subsequently, they made use of a random walk Metropolis Hastings algorithm with a Gaussian proposal density centered at the current parameter value to generate samples from analytically intractable posterior distribution of $\beta_p$. 

In the aforementioned approach, the acceptance probability depends on the choice of the value of $p$, hence the fine tuning of parameters like proposal step size is necessary to obtain the appropriate acceptance rates for each $p$. \cite{Kozumi-Kobayashi-2011} overcame this limitation and showed that Gibbs sampling can be incorporated with AL density being represented as a mixture of normal and exponential distributions. Consider the linear model from \autoref{quant_lin_model}, where $\epsilon_i \sim$ ALD$(0,\sigma,p)$, then this model can be written as
\begin{equation} \label{quant_lin_ald_model}
y_i = x_i^T {\beta}_p + \theta v_i + \tau \hspace{0.1cm}\sqrt[]{\sigma v_i} u_i, \hspace{1cm} i = 1,...,n,
\end{equation}
where, $v_i$ and $u_i$ are mutually independent, with $u_i \sim $N$(0,1), v_i \sim \varepsilon(1/\sigma)$ and $\varepsilon$
represents exponential distribution. The $\theta$ and $\tau$ constants in the \autoref{quant_lin_ald_model} are given by 
\begin{equation} \label{theta_tau}
\theta = \frac{1-2p}{p(1-p)} \quad and \quad \tau = \sqrt[]{\frac{2}{p(1-p)}}
\end{equation}
Consequently, a Gibbs sampling algorithm based on standard normal distribution can be implemented effectively. Currently, \texttt{Brq} \cite{Alhamzawi-2017} and \texttt{bayesQR} \cite{Benoit-et-al-2017} are two R packages which provide Gibbs sampler for Bayesian quantile regression. We are exploiting the same technique to derive a Gibbs sampler for Bayesian quantile regression in longitudinal count data models. 



\section{Bayesian Quantile Regression for Longitudinal Count Data} 

\label{Section3} 


\subsection{Model Framework}

Let $(x_{ij}^{T},y_{ij})$ be the repeated measurements data, for $(i=1,...,N; j=1,...,n_i)$, where $y_{ij}$ denotes the response measured for the $i$th subject at $j$th time and $x_{ij}^{T}$  is $1 \times k$ row vector of a known design matrix $X_i$. The response variables are counts and assumed to be generated via Poisson counting process. Poisson log-linear models have been typically used for such analyses and a general formulation is given by:
\begin{gather*} 
Y_{ij} \sim \mathrm{Poisson}(\mu_{ij}), 
\\ \mu_{ij} = \hspace{1mm} \exp \{x_{ij}^{T}\beta + s_{ij}^{T}\alpha_i\}
\end{gather*}
The longitudinal count data model can be expressed in the latent variable formulation as,
\begin{equation} \label{long_count_latent_model}
z_{ij} = x_{ij}^{T}\beta + s_{ij}^{T}\alpha_i + {\epsilon}_{ij}, \hspace{1cm} i = 1,...,N, j = 1,...,n_i,
\end{equation}
where $z_{ij}$ denotes $j$th latent variable for $i$th individual, $\beta$ is $k \times 1$ vector of unknown fixed-effects parameters, $s_{ij}^{T}$ is a $1 \times l$ row vector of covariates possessing random effects, $\alpha_i$ is $l \times 1$ vector of unknown random-effects parameters and $\epsilon_{ij}$ are the errors which are assumed to be $i.i.d.$ and  following ALD$(0,\sigma,p)$. Thus, the $p^{\text{th}}$ quantile level linear mixed quantile functions can be formulated in terms of a latent variable $z_{ij}$ as given by
\begin{equation} \label{quant_latent_fun}
Q_{z_{ij}}(p |x_{ij},s_{ij}) = x_{ij}^{T}\beta + s_{ij}^{T}\alpha_i
\end{equation}
Latent variables in the above expression remain unobserved and only count variables are observed. \cite{Koenker-Basset-1978} provided sufficient conditions for asymptotic inference of the parameters of $Q_{y_i}(p |x_i)$ stated in \autoref{quant_fun}. Among them, the continuity and positive support of the pdf of $y$ conditional on $x$ are the conditions of concern. Here, $Y$ is generated through a counting process and its support is a set of nonnegative integers which does not satisfy both of the conditions specified. \cite{Machado-Silva-2005} dealt with this problem by jittering the $Y$ variable with uniform noise to produce continuous variables. These jittered responses, $y_{ij}^{*} = y_{ij} + u_{ij}$, where $u_{ij} \sim$ unif$(0,1)$, have continuous real valued quantiles and can be modeled as:
\begin{equation} \label{quant_real_fun}
Q_{y_{ij}^{*}}(p |x_{ij},s_{ij}) = p + \exp \{x_{ij}^{T}\beta + s_{ij}^{T}\alpha_i \}
\end{equation}
Models \eqref{quant_latent_fun} and \eqref{quant_real_fun} are equivalent as we can obtain $z_{ij}$ from $y_{ij}^{*}$ through a monotonic transformation, 
\begin{equation} \label{z_y_transform}
z_{ij} = \left\{\begin{matrix}
\ln(y_{ij}^{*}-p), & \enskip \text{for} \enskip y_{ij}^{*} > p 
\\
\ln(\zeta)
 & \enskip \text{for} \enskip y_{ij}^{*} \leq p
\end{matrix}\right.
\end{equation}
where, $\zeta$ is a suitably small positive number. I take $\zeta = 10^{-5}$ throughout this work. We can rewrite the \autoref{quant_real_fun} as
\begin{equation*}
Q_{z_{ij}}(p |x_{ij},s_{ij}) = \ln\{Q_{y_{ij}^{*}}(p |x_{ij},s_{ij}) - p\}
\end{equation*}
To take into account the variation introduced by contamination of the count responses with uniformly distributed jittering variables, \cite{Machado-Silva-2005} applied their model to $M$ jittered data sets $Y+U^{(h)}$, for $h = 1,...,M$. They estimated the parameters of underlying regression as follows:
\begin{equation*} \label{l}
\hat{\beta}_p = \sum_{h=1} ^M \overset{\sim}{\beta}_p^{(h)} 
\end{equation*}
where, $\overset{\sim}{\beta}_p^{(h)}$ is the estimate obtained for a simulation on jittered response data $Y+U^{(h)}$. Instead of deriving expression for asymptotic covariance matrix, we can calculate the credible interval of average estimate by averaging the credible intervals of $M$ estimates, as given in \cite{Lee-Neocleous-2010}. The estimated conditional quantile function is specified as:
\begin{equation*} 
\hat{Q}_{y_{ij}}(p |x_{ij},s_{ij}) = \lceil \hat{Q}_{y_{ij}^{*}}(p |x_{ij},s_{ij}) - 1 \rceil
= \lceil p + \exp \{x_{ij}^{T}\beta + s_{ij}^{T}\alpha_i\} - 1 \rceil,
\end{equation*}
where, $\lceil . \rceil$ is the ceiling function. Now, we can give the complete data density of latent variables $z_{ij}$, conditional on $\alpha_i$, for $i=1,...,N$ and $j=1,...,n_i$, with the assumption that they are identically and independently distributed according to the ALD$(x_{ij}^{T}\beta + s_{ij}^{T}\alpha_i,\sigma,p)$.
\begin{equation*} 
f_z(z_{ij}|\beta,\alpha_i,\sigma) = \frac{p(1-p)}{\sigma} \exp \left \{ -\rho_p \left ( \frac{z_{ij} - x_{ij}^T\beta - s_{ij}^T\alpha_i}{\sigma} \right) \right \}
\end{equation*}
In our case, the conditional quantile to be estimated, $p$, is fixed and known beforehand. The random effects introduced here induce autocorrelation among the observations on the same subject. We assume that $\alpha_i \overset{iid}\sim f_\alpha (\alpha_i|\Sigma)$ and $\epsilon_{ij}$ terms are independent and further $\alpha_i$ and $\epsilon_{ij}$ are independent of each other. 

Let $z_i = (z_{i1},...,z_{in_i})$ and $f_z(z_i|\beta,\alpha_i,\sigma) = \prod_{j=1}^{n_i} f_z(z_{ij}|\beta,\alpha_i,\sigma)$ be the conditional density of latent response of $i^{\text{th}}$ subject conditional on the random effect $\alpha_i$. The complete latent data density of $(z_i,\alpha_i)$, for $i = 1,...,N,$ is given by
\begin{equation*} 
f(z_i,\alpha_i|\beta,\sigma,\Sigma) =f_z(z_i|\beta,\alpha_i,\sigma) f_\alpha(\alpha_i|\Sigma).
\end{equation*}
Let $\mathbf{z} = (z_1,...,z_N)$ and $\bm{\alpha} = (\alpha_1,...,\alpha_N)$, the joint latent data density of $(z,\alpha)$ for N individuals is as follows:
\begin{equation} \label{joint_latent_density}
f(\mathbf{z},\bm{\alpha}|\beta,\sigma,\Sigma) =\prod_{i=1}^{N}f_z(z_i|\beta,\alpha_i,\sigma) f_\alpha(\alpha_i|\Sigma).
\end{equation}

\subsection{The Hierarchical Model}

After exploiting the normal-exponential mixture representation representing AL distribution and assuming the priors $\beta \sim \pi(\beta), \sigma \sim \pi(\sigma), \Sigma \sim \pi(\Sigma) $, we get the following hierarchical Bayesian quantile regression model:
\begin{eqnarray} \label{hierar_bqr_model}
 z_{ij} & = & x_{ij}^{T}\beta + s_{ij}^{T}\alpha_i + \theta v_{ij} + \tau \hspace{0.1cm}\sqrt[]{\sigma v_{ij}} u_{ij}, \nonumber \\
 v_{ij} & \sim & \varepsilon(1/\sigma),  \quad u_{ij} \hspace{1mm} \sim \hspace{1mm} N(0,1), \nonumber \\
 \beta & \sim & \pi(\beta), \quad \sigma \hspace{1mm} \sim \hspace{1mm} \pi(\sigma) \\
 \alpha_i|\Sigma & \sim & f_\alpha (\alpha_i|\Sigma), \quad
 \Sigma \hspace{1mm} \sim \hspace{1mm} \pi(\Sigma). \nonumber
\end{eqnarray}
We can obtain the posterior distributions of parameters using the Bayes theorem in \autoref{joint_latent_density} as follows:
\begin{equation*} 
f(\beta,\sigma,\Sigma,\alpha|z) \enskip \propto \enskip f(z,\alpha|\beta,\sigma,\Sigma) \hspace{0.1cm} \pi(\beta) \hspace{0.1cm} \pi(\sigma) \hspace{0.1cm} \pi(\Sigma),
\end{equation*}
Usually we are interested only in the inference regarding fixed-effects parameters $\beta$ which can be accomplished by obtaining the marginal distribution by integrating out the $\sigma, \Sigma, \alpha$ parameters from $f(\beta,\sigma,\Sigma,\alpha|z)$ as given below
\begin{equation*} 
f(\beta|z) \hspace{0.1cm} = \hspace{0.1cm} \int ... \int f(z,\alpha|\beta,\sigma,\Sigma)\hspace{0.1cm} d \alpha \hspace{0.1cm} d \sigma \hspace{0.1cm} d \Sigma.
\end{equation*}
The next step in estimation is to choose appropriate prior distributions for each parameter. However, one needs to be cautious while selecting priors else issues can arise which could hamper the estimation and thereby inference process \cite{Kinney-Dunson-2007}. 

We have taken the normal prior, N$(0,\phi^2I)$, for the mutually indepedent random-effects parameters. As pointed out by \cite{Koenker-2004}, prior on $\alpha_i$ has a penalty interpretation on the quantile loss function, and in my case, the normal prior implies $l_2$ penalization. $l_1$ penalization can be imposed by choosing an AL distribution as a prior \cite{Geraci-Bottai-2007}. One could set the prior to be N$(0,\Sigma)$, but inference in such scenario becomes computationally expensive owing to the additional parameters to be estimated. Finally, we consider a hyper prior for $\phi^2$, an inverse gamma (IG) distribution with shape parameter $b_1$ and scale parameter $b_2$.
\begin{eqnarray*}
\alpha_i \enskip \sim \enskip \text{N} (0,\phi^2I), \quad  
\phi^2 \enskip \sim \enskip \text{IG}(b_1,b_2).
\end{eqnarray*}
For the fixed-effects $\beta$, the conventional choice of normal prior with zero mean leads to the ridge estimator. \cite{Griffin-Brown-2010} showed that if differences in the size of fixed effects are large then the normal prior performs poorly. Instead we consider the Laplace prior for the fixed-effects, which takes the form as follows:
\begin{equation*} 
\pi(\beta|\lambda) =\prod_{t=1}^{k} \hspace{0.1cm} \frac{\lambda}{2} \hspace{0.1cm} \exp(-\lambda \hspace{0.1cm} |\beta_t|), \qquad \lambda \geq 0.
\end{equation*}
Laplace prior is a generalization of the ridge prior, which is equivalent to the Lasso model \cite{Tibshirani-1996,Bae-Mallick-2004}. The above prior can be written as given by \cite{Andrews-Mallows-1974},
\begin{gather*} 
\prod_{t=1}^{k} \frac{\lambda}{2} \exp(-\lambda | \beta_t|) \hspace{1mm} = \hspace{1mm} \prod_{t=1}^{k} \int _0 ^\infty \text{N}(\beta_t;0,g_{t}^2) \hspace{0.1cm} \varepsilon (g_t^2;\frac{\lambda^2}{2}) \hspace{0.1cm} d g_t, \\
g_t^2 \hspace{1mm} \sim \hspace{1mm} \varepsilon(\frac{\lambda^2}{2}), \quad
\lambda^2 \hspace{1mm} \sim \hspace{1mm} \text{Gamma}(a_1,a_2).
\end{gather*}
Further the scale parameter $\sigma$ is assigned a conjugate inverse gamma prior, $\sigma  \sim$ IG$(c_1,c_2)$, which allows it to be dynamically updated at each iteration in the Gibbs sampling algorithm. It removes the need to tune $\sigma$ to obtain good acceptance rates such as in MCMC sampling using a Metropolis-Hastings algorithm. 

Efficient values of parameters in an inverse-gamma prior have been debated a lot in the recent literature. The ordinary choice of shape and rate parameters to be each equal to 0.01 is widely criticized as it assigns negligible weight to small values of parameter on which prior beliefs are made. Following \cite{Gelman-2006}, we specify a flat prior on $\sigma$ and $\phi^2$ with the shape parameter set to $-0.5$ and rate parameter set to $0$. It is nothing but an improper IG distribution which retains the conjugacy property of a proper IG distribution and at the same time remains vague in nature.

The joint posterior distribution of all parameters given the latent variable $z$ is formulated as given below:
\begin{eqnarray*}
f(\bm{\beta},\bm{\alpha},\mathbf{v},\sigma,\phi^2,\mathbf{g^2},\lambda^2|\mathbf{z}) & \propto & f(\mathbf{z}| \bm{\beta},\bm{\alpha},\mathbf{v},\sigma) \hspace{0.1cm} f(\mathbf{v}|\sigma) \hspace{0.1cm} f(\sigma)  \\
 & \times & f(\bm{\beta}|\mathbf{g^2}) \hspace{0.1cm} f(\mathbf{g^2}|\lambda^2) \hspace{0.1cm} f(\lambda^2)\\ 
 & \times & f(\bm{\alpha}|\phi^2) \hspace{0.1cm} f(\phi^2),
\end{eqnarray*}
where, $\mathbf{z} = (z_{11},...,z_{N,n_N})$, $\mathbf{v} = (v_{11},...,v_{N,n_N})$, $\bm{\beta} = (\beta_1,...,\beta_k)$, $\bm{\alpha} = (\alpha_1,...,\alpha_N)$ and $\mathbf{g^2} = (g_1^2,..., g_k^2)$. After substituting the probability distribution functions in the above equation it produces the following expression,
\begin{align*}
& f(\bm{\beta},\bm{\alpha},\mathbf{v},\sigma,\phi^2,\mathbf{g^2},\lambda^2|\mathbf{z}) \\
& \hspace{2mm} \propto \hspace{1mm} \left \{ \prod_{i=1}^{N} \left \{ \prod_{j=1}^{n_i} ( 2 \pi \tau^2 \sigma v_{ij})^{-\frac{1}{2}} \exp \left [ -\frac{1}{2 \tau^2 \sigma v_{ij}} (z_{ij}-x_{ij}^{T}\beta - s_{ij}^{T}\alpha_i-\theta v_{ij})^2 \right] \right. \right. \\ 
& \hspace{12mm} \left. \left. \exp(-\frac{v_{ij}}{\sigma})\right \} (2 \pi \phi^2)^{-\frac{1}{2}} \exp \left [ -\frac{1}{2} \frac{\alpha_i^{T}\alpha_i}{\phi^2} \right ] \right \}  \\ 
& \hspace{7mm} \times  (\sigma)^{-(c_1+1)} \exp \left [-\frac{c_2}{\sigma} \right ] \\ 
& \hspace{7mm} \times (\phi^2)^{-(b_1+1)} \exp \left [-\frac{b_2}{\phi^2} \right ]  \\ 
& \hspace{7mm} \times (2 \pi)^{-\frac{k}{2}}|D_{g^2}^{-1}|\exp \left [ -\frac{1}{2} \beta^{T} D_{g^2}^{-1} \beta \right ] \\
& \hspace{7mm} \times \left \{ \prod_{h=1}^{k} \exp \left [ -\frac{\lambda^2}{2} g_h^2 \right ] \right \} \\
& \hspace{7mm} \times  (\lambda^2)^{a_1-1} \exp \left [-a_2 \lambda^2 \right ].
\end{align*}
Where, $\theta$ and $\tau$ are the constants mentioned in the \autoref{theta_tau} and $D_{g^2}$ is $k \times k$ diagonal matrix with $(g_1^2,...,g_k^2)$ being the diagonal elements. 

This joint posterior distribution does not possess a tractable form and Markov chain Monte Carlo simulation methods are used to carry out the Bayesian inference and eventually obtain the estimates of the parameters. Rather than sampling each component individually we consider within block sampling for $(\beta,\alpha)$ similar to \cite{Luo-et-al-2012}. In their model, they sampled both fixed and random effects from their respective conditional posterior distributions in a linear mixed-effects model. This block sampling of parameters accounts for possible correlation among the components of $\beta$ and $\alpha$. 

In within block sampling, we sample $\beta$ conditional on $\alpha$ from an updated normal distribution and similarly $\alpha$ is sampled conditional on $\beta$ from an another updated normal distribution. The latent variable $v$ is sampled from an updated generalized inverse Gaussian (GIG) distribution. The latent data variable $z$ is generated using the expression in \autoref{z_y_transform}. In this process, jittering variables from uniform distribution are sampled at each iteration and utilized to obtain the latent variable $z$ whose conditional quantiles are continuous and hence can be modeled by a linear combination of covariates. The uniform jittering variables possess no practical importance and at each iteration new values are generated discarding the old ones. The parameters $\sigma$ and $\phi^2$ are sampled from updated IG distributions. On the other hand, $g_t^2$ is sampled from an updated GIG distribution and $\lambda^2$ is sampled from an updated gamma distribution. A detailed description of the Gibbs sampling algorithm is given below.

\subsection*{Algorithm}
\noindent\rule{\textwidth}{0.4pt}
\begin{enumerate}
\item Sample $z|y$ according to the \autoref{z_y_transform} 
\begin{align*}
& z_{ij} = \left\{\begin{matrix}
\ln(y_{ij}^{*}-p), & \quad \text{for} \enskip y_{ij}^{*} > p 
\\
\ln(\zeta)
 & \quad \text{for} \enskip y_{ij}^{*} \leq p
\end{matrix}\right. \\
& \text{where, } y_{ij}^{*} = y_{ij} + u_{ij} \text{ and } u_{ij} \text{ are sampled from } \mathrm{unif}(0,1); \\
& \hspace{14mm} \zeta = 10^{-5} \text{ and } p \text{ is the quantile level.}
\end{align*}
\item Sample $v_{ij}|z,\beta,\alpha_i,\sigma \sim \mathrm{GIG}(\frac{1}{2},\rho_1,\rho_2)$ for for $i=1,...,N$ and $j=1,...,n_i$, where, 
\begin{align*}
& \rho_1 = \frac{(z_{ij}-x_{ij}^T \beta - s_{ij}^T \alpha_i)^2}{\tau^2 \sigma} \quad \text{and} \quad \rho_2 = \frac{\theta^2}{\tau^2 \sigma} + \frac{2}{\sigma}.
\end{align*}
\item Sample $\sigma|z,v,\beta,\alpha,\phi^2 \sim \mathrm{IG}(\tilde{c}_1,\tilde{c}_2)$, where,
\begin{align*}
& \tilde{c}_1 = \frac{n_i N}{2} + c_1 \quad \text{and} \quad \tilde{c}_2 = \frac{1}{2 \tau^2} \sum_{i=1}^N \sum_{j=1}^{n_i} \frac{(z_{ij}-x_{ij}^T\beta - s_{ij}^T\alpha_i - \theta v_{ij})^2}{v_{ij}} + c_2.
\end{align*}
\item Sample $\beta$ conditional upon $\alpha$ from the distribution $\beta|z,v,\alpha,\sigma,g^2 \sim \mathrm{N}(\tilde{\beta},\tilde{B})$, where,
\begin{align*}
& \tilde{B}^{-1} =  \left ( \sum_{i=1}^N \frac{x_i^{T} (D_{v_i}^2)^{-1} x_i}{\tau^{2 n_i}} + D_{g^2}^{-1} \right ) \enskip \text{and} \enskip \tilde{\beta}  =  \tilde{B} \left ( \sum_{i=1}^N \frac{x_i^{T}(D_{v_i}^2)^{-1}(z_i-s_i^{T} \alpha_i - \theta v_i)}{\tau^{2 n_i}} \right ) \\
& \text{where, } \hspace{0.1cm}  D_{v_i} \text{ is the } \hspace{0.05cm} \text{diag}\hspace{0.05cm}(\hspace{0.1cm}\sqrt[]{\sigma v_{i1}}\hspace{0.1cm},...,\hspace{0.1cm}\sqrt[]{\sigma v_{i n_i}}\hspace{0.1cm}) \text{ matrix.} 
\end{align*}
\item Sample $g_h^2|\beta_h \sim \mathrm{GIG}(\frac{1}{2},\rho_3,\rho_4)$ for $h=1,...,k$, where, $\rho_3 = \beta_h^2$ and $\rho_4 = \lambda^2$.
\item Sample $\lambda^2|g_h^2 \sim \mathrm{Gamma}(\tilde{a}_1,\tilde{a}_2)$, where, $\tilde{a}_1 = k + a_1$ and $\tilde{a}_2 = \sum_{h=1}^k \frac{g_h^2}{2} + a_2$.
\item Sample $\alpha$ conditional on $\beta$ from the distribution $\alpha_i|z,v,\beta,\sigma,\phi^2 \sim \mathrm{N}(\tilde{a},\tilde{A})$ for $i = 1,...,N$, where,
\begin{align*}
& \tilde{A}^{-1} = \left ( \frac{s_i^{T} (D_{v_i}^2)^{-1} s_i}{\tau^{2 n_i}} + \frac{1}{\phi^2} I_l \right ) \quad \text{and} \quad \tilde{a} = \tilde{A} \left ( \frac{s_i^{T} (D_{v_i}^2)^{-1} (z_i-x_i^{T} \beta - \theta v_i)}{\tau^{2 n_i}}\right ).
\end{align*}
\item Sample $\phi^2|\alpha \sim \mathrm{IG}(\tilde{b}_1,\tilde{b}_2)$, where, $\tilde{b}_1 = \frac{n_i N}{2} + b_1$ and $\tilde{b}_2 = \sum_{i=1}^{n_i} \frac{\alpha_i^T\alpha_i}{2} + b_2$.
\end{enumerate}
\noindent\rule{\textwidth}{0.4pt}

\section{Empirical Illustrations} 

\label{Section4} 


\subsection{Simulation Studies}


\subsubsection{Random Intercept Model}
\begin{figure}[H]
\begin{subfigure}[h]{0.5\textwidth}
\centering
\includegraphics[width=0.8\textwidth]{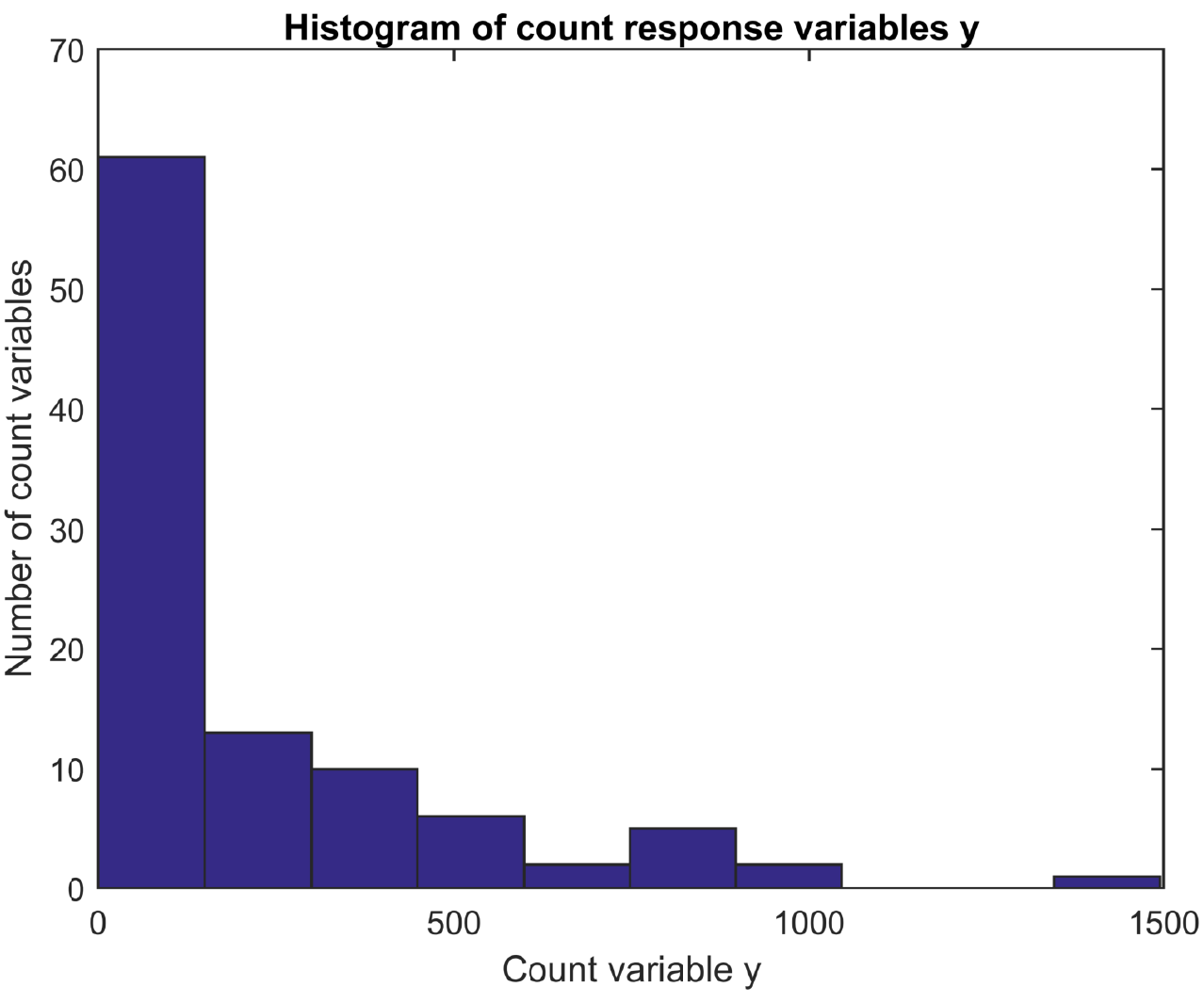}
\caption{Histogram of y}
\label{hist_responses_sim_1}
\end{subfigure}
\begin{subfigure}[h]{0.5\textwidth}
\centering
\includegraphics[width=0.8\textwidth]{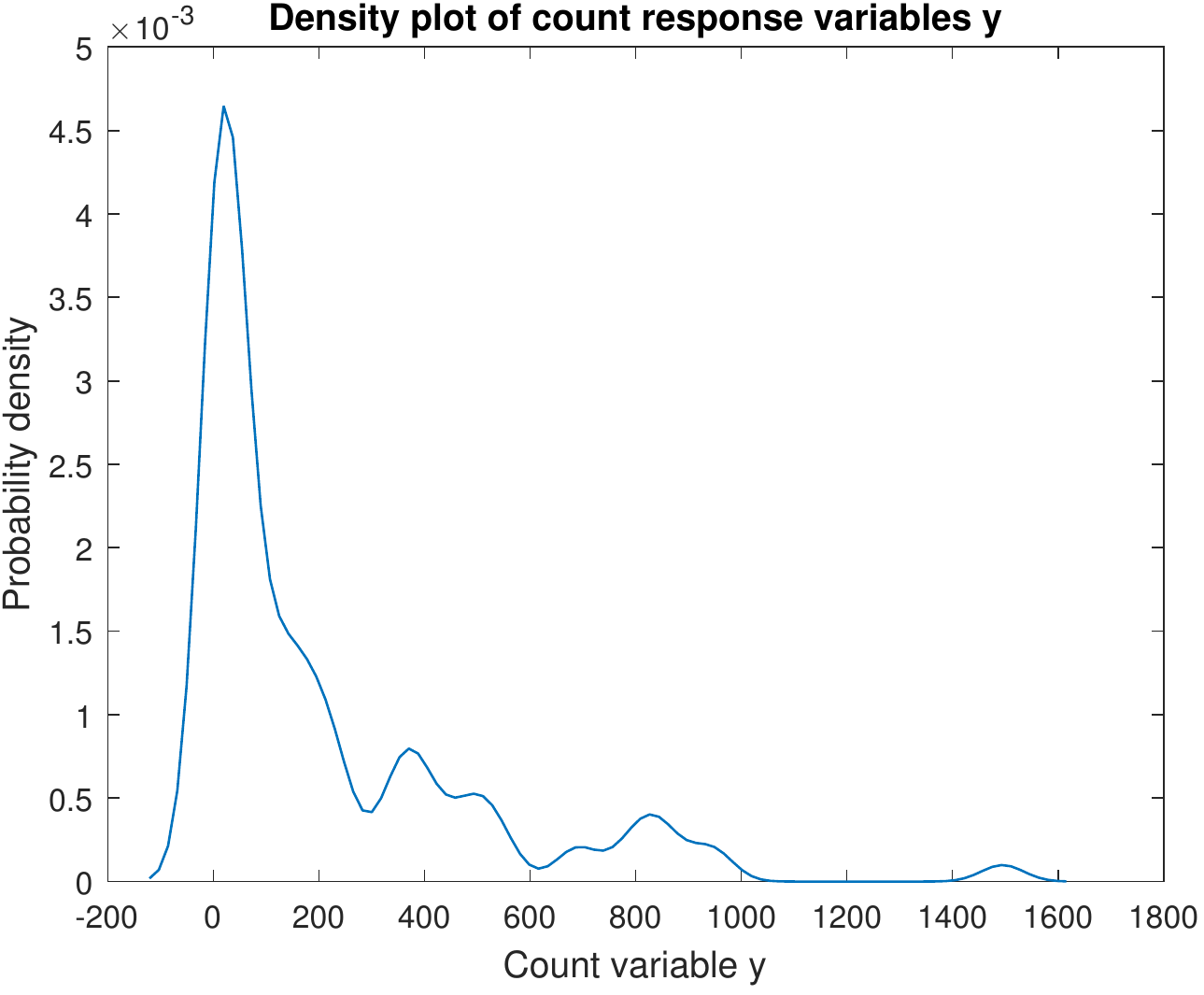}
\caption{Probability density plot of y}
\label{density_responses_sim_1}
\end{subfigure}
\caption{Count response data plots for simulation study 1}
\label{responses_sim_1}
\end{figure}

We considered the following simple linear mixed-effects model for latent variable formulation
\begin{equation} \label{sim_1_model}
z_{ij} = x_{1ij}\beta_1 + x_{2ij}\beta_2 + x_{3ij}\beta_3 + \alpha_i + {\epsilon}_{ij}, \hspace{1cm} i = 1,...,N, j = 1,...,n_i,
\end{equation}	
where, number of subjects $(N)$ were 20 and for each subject number of repeated measurements $(n_i)$ were equal to 5. $x_{1ij}, x_{2ij}$ and $x_{3ij}$ were sampled independently from a uniform distribution on the interval $[0,1]$ and $(\beta_1,\beta_2,\beta_3) = (1,3,5)$. A random intercept effect $\alpha_i \sim$ N$(0,1)$ was considered. As we don't consider the fixed intercepts, the random intercepts account for them as well as the subject specific deviations. In this experiment, the counts $y_{11},...,y_{N n_i}$ were generated according to Poisson process. Hence, the counts were Poisson random variables with conditional mean given by
\begin{equation} \label{sim_1_mu}
\mu_{ij} = \exp \{ x_{1ij}\beta_1 + x_{2ij}\beta_2 + x_{3ij}\beta_3 + \alpha_i\}, \hspace{1cm} i = 1,...,20, j = 1,...,5.
\end{equation}	
The histogram of count responses along with the smoothed kernel density plot is given in \autoref{responses_sim_1}. We then carried out Bayesian quantile regression at three different quantile levels, $p=(0.25,0.50,0.75)$. We assumed weak prior information and assigned IG$(-0.5,0)$ prior for $\sigma$ as well as $\phi^2$ and considered Gamma$(0.01,0.01)$ prior on $\lambda^2$. To average out the noise added because of jittering, we implemented the average jittering estimator for each quartile with $M=20$ independent jittered data. According to \cite{Machado-Silva-2005}, even with a moderate number of repetitions (say $M=10$) we can achieve high precision in average estimator as compared to a single posterior estimate. The Bayesian quantile regression using Gibbs sampling algorithm proposed above was run for r = 10000 iterations after initial 2000 iterations were discarded for burn-in on each jittered data. 

The results for each quartile are presented in \autoref{sim_study_1}. We can see that the average posterior mean estimates for $\beta$ are close to the true parameter values. Standard deviations for these average estimators were calculated using the pooled variance technique given below. Average credible interval thus obtained was shorter than the credible interval of posterior mean for each jittered data revealing the improvement in precision of the average estimator.
\begin{gather} \label{pool_var}
\mathrm{Var}(\hat{\beta}_p) = \left ( 1- \frac{1}{r} \right ) W + \frac{1}{r} B \\
\text{where,} \enskip
W = \frac{1}{M} \sum_{h=1}^M \mathrm{Var}(\overset{\sim}{\beta}_p^{(h)}) \quad \text{and} \quad B = \frac{r}{M-1} \sum_{h=1}^M (\overset{\sim}{\beta}_p^{(h)} - \hat{\beta}_p)^2 \nonumber
\end{gather}

\begin{figure}[H]
\includegraphics[width=\textwidth]{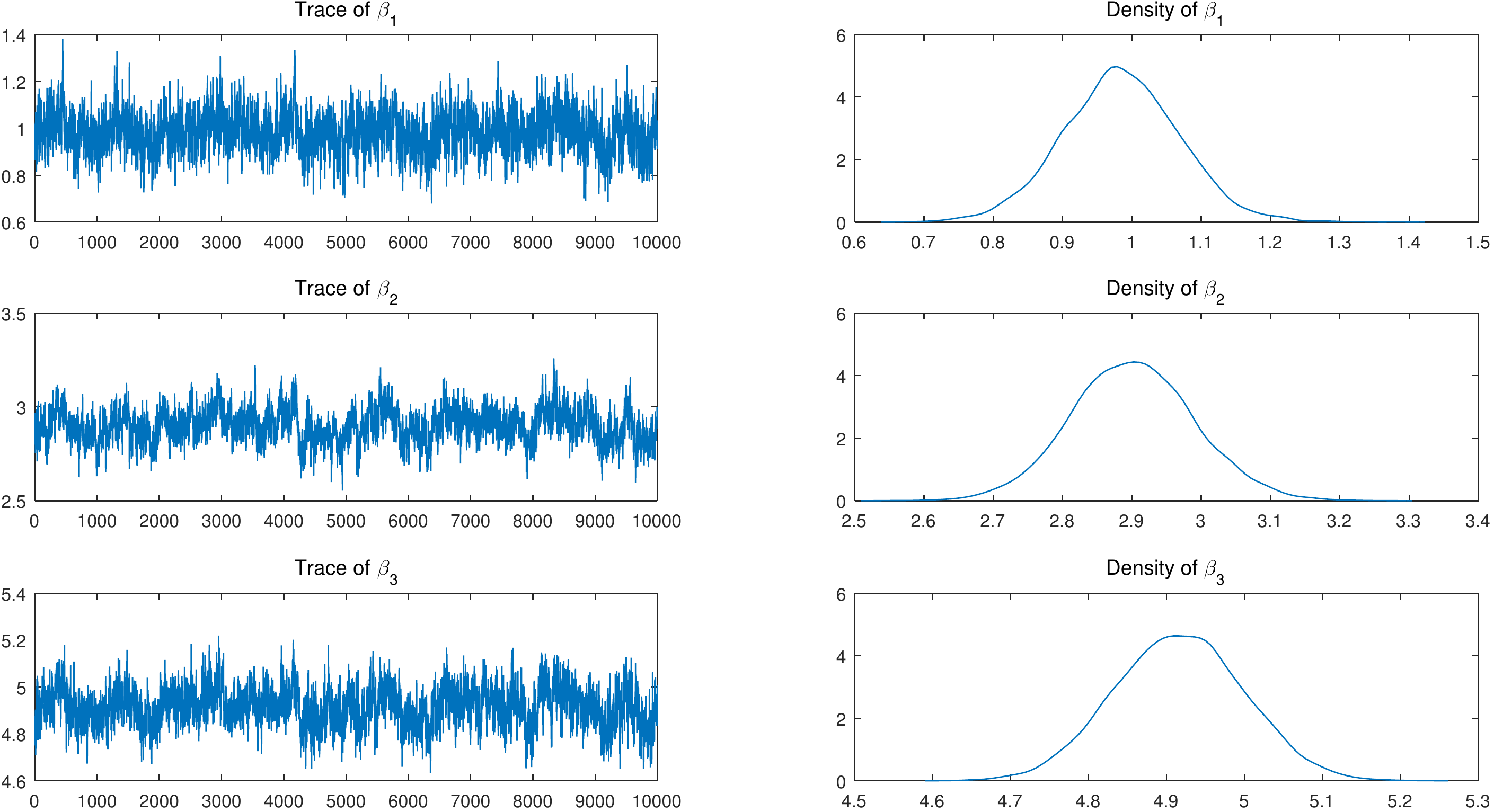}
\centering
\caption{Trace plots and kernel density plots of $\beta_1, \beta_2$ and $\beta_3$ at 0.25 quantile based on 10,000 iterations on one of the jittered dataset in random intercept model}
\label{trace_density_sim_1_quartile_1_beta}
\end{figure}

\begin{figure}[H]
\includegraphics[width=\textwidth]{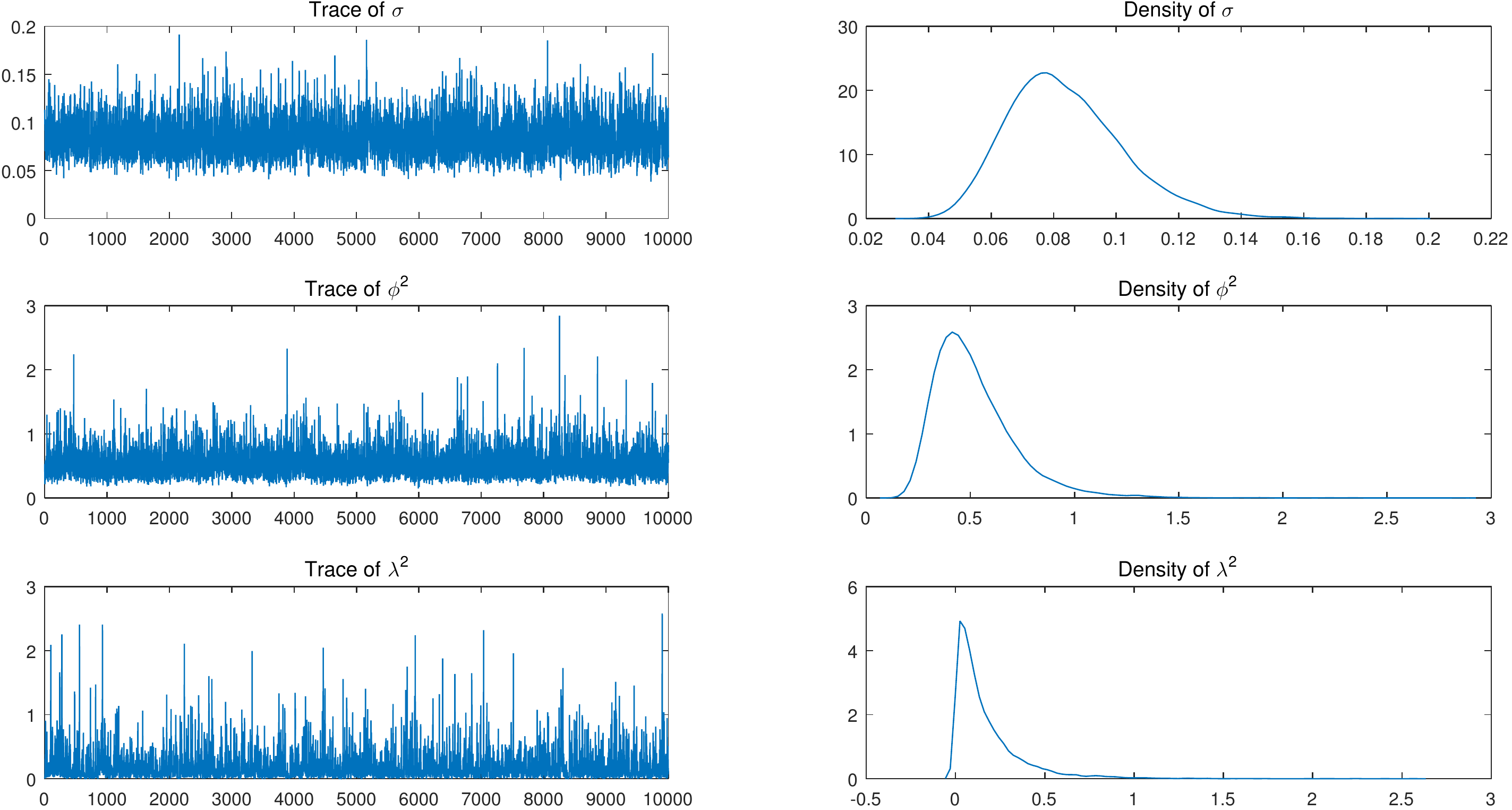}
\centering
\caption{Trace plots and kernel density plots of $\sigma, \phi^2$ and $\lambda^2$ at 0.25 quantile based on 10,000 iterations on one of the jittered dataset in random intercept model}
\label{trace_density_sim_1_quartile_1_hyper}
\end{figure}

Trace plots and kernel density plots for $\beta$ and other hyperparameters $(\sigma, \phi^2, \lambda^2)$ at $p = 0.25$ quantile regression are shown in \autoref{trace_density_sim_1_quartile_1_beta} and \autoref{trace_density_sim_1_quartile_1_hyper} respectively. 
Convergence of chains for each parameter is observed through these trace plots which graphically demonstrate the appealing sampling achieved by our block sampling algorithm.
Similar graphs were obtained for $p = (0.50,0.75)$ quantile regressions and can be made available on request. 


To compare the Bayesian quantile regression models with the standard classical mean regression model, we performed random-effects Poisson regression (REPR) and results are presented in \autoref{mean_reg_sim_1}. We can see that coefficient values estimated are close to the true values and they are statistically significant as well. Our model gives average posterior means of the parameters close to their true values with slightly wider credible intervals in comparison to REPR model. Next, we calculated the negative log-likelihood (NLL) to compare the Bayesian quantile regression models with the classical model due to lack of a common information criterion across frequentist and Bayesian setting. The NLL for the REPR model was 398.5 and for our model the values for $p=(0.25, 0.50, 0.75)$ were 587.7, 390.7 and 417.7 respectively. This suggests that $p=0.50$ Bayesian quantile regression model fits the data better than any other model. 

Furthermore, the model selection criterion such as deviance information criterion (DIC) \cite{Spiegelhalter-et-al-2002,Celeux-et-al-2006} proves helpful in choosing a value of $p$ which is most consistent with the data. The DIC values for $1^\text{st}, 2^\text{nd}$ and $3^\text{rd}$ quartiles are 1878.8, 1142.7 and 1176.5 respectively. Hence, according to the DIC, median model provides the best fit among these three quantile regression models.


\begin{table}[H]
\centering
\caption{Average posterior means, standard deviations and 95\% average credible intervals at $p=0.25, 0.50, 0.75$ quantile levels}
\label{sim_study_1}
\begin{tabular}{@{}lcccccc@{}}
\toprule
Quantile & Par & True $\beta$ & Avg Post Mean & SD & Avg $2.5\%$ & Avg $97.5\%$ \\
\midrule
$p=0.25$ 
& $\beta_1$ & 1 & 0.9981 & 0.1421 & 0.7039 & 1.2651 \\ 
& $\beta_2$ & 3 & 2.8788 & 0.1454 & 2.5829 & 3.1532 \\
& $\beta_3$ & 5 & 4.9470 & 0.1288 & 4.6892 & 5.1974 \\ \midrule
$p=0.50$ 
& $\beta_1$ & 1 & 0.9856 & 0.0985 & 0.7899 & 1.1763 \\ 
& $\beta_2$ & 3 & 2.8957 & 0.1001 & 2.6947 & 3.0882 \\
& $\beta_3$ & 5 & 4.9506 & 0.0915 & 4.7694 & 5.1293 \\ \midrule
$p=0.75$ 
& $\beta_1$ & 1 & 0.9708 & 0.0830 & 0.8064 & 1.1334 \\ 
& $\beta_2$ & 3 & 2.8790 & 0.0864 & 2.7101 & 3.0483 \\
& $\beta_3$ & 5 & 4.8970 & 0.0840 & 4.7332 & 5.0606 \\ 
\bottomrule
\end{tabular}
\end{table}

\begin{table}[H]
\centering
\caption{Random Effects Poisson Regression (REPR) for simulation study 1}
\label{mean_reg_sim_1}
\begin{tabular}{cccccc}
\toprule
Par & True $\beta$ & Coef  & Std Err & \multicolumn{2}{c}{95\% Conf Int} \\ \hline
$\beta_1$ & 1  & 0.9360 & 0.0397  & 0.8581  & 1.0139  \\
$\beta_2$ & 3  & 2.9308 & 0.0416  & 2.8493  & 3.0124  \\
$\beta_3$ & 5  & 4.8937 & 0.0474  & 4.8009  & 4.9865  \\ 
\bottomrule
\end{tabular}
\end{table}

\subsubsection{Random Intercept and Slope Model}
\begin{figure}[h]
\begin{subfigure}[h]{0.5\textwidth}
\centering
\includegraphics[width=0.8\textwidth]{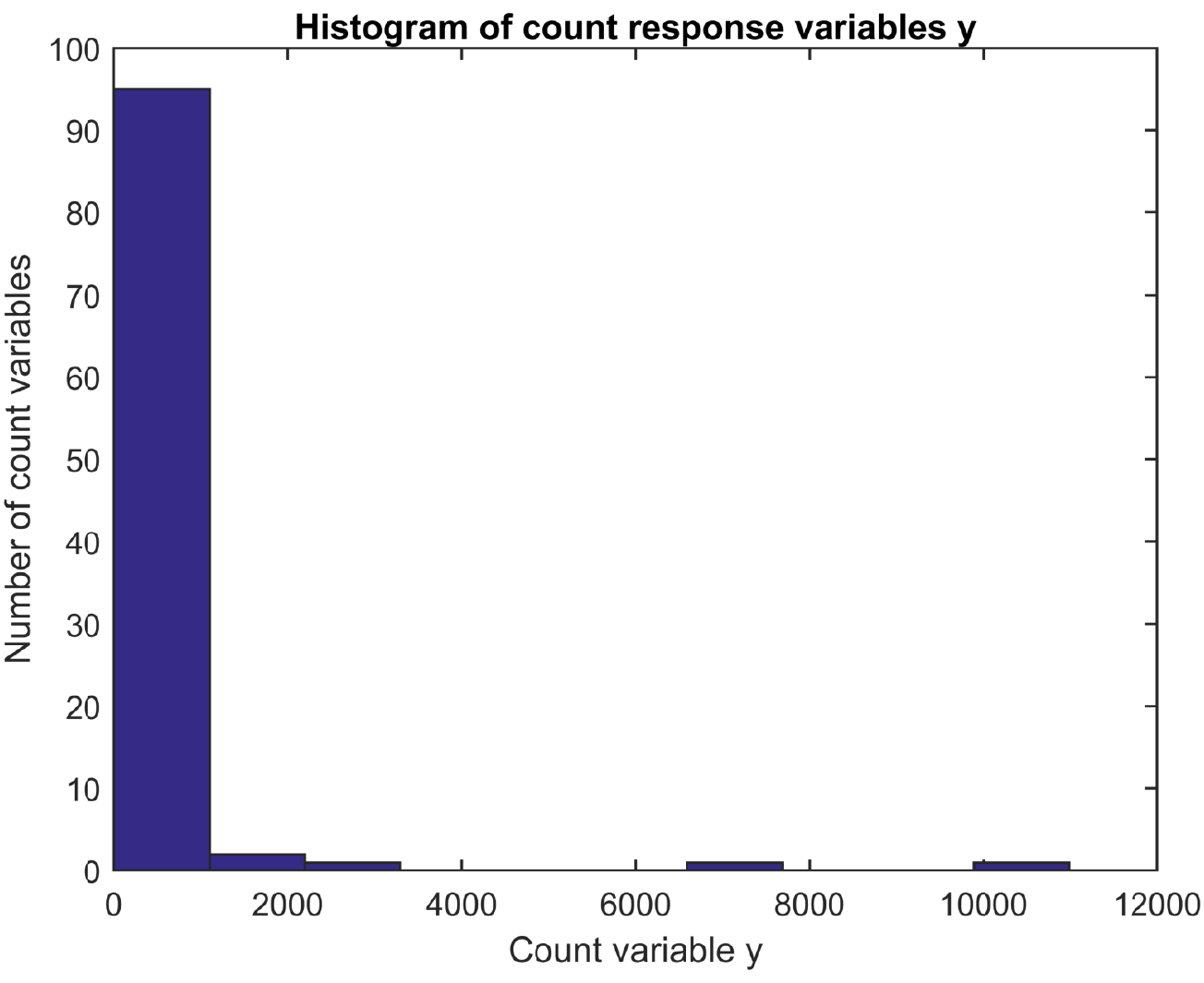}
\caption{Histogram of y}
\label{hist_responses_sim_2}
\end{subfigure}
\begin{subfigure}[h]{0.5\textwidth}
\centering
\includegraphics[width=0.8\textwidth]{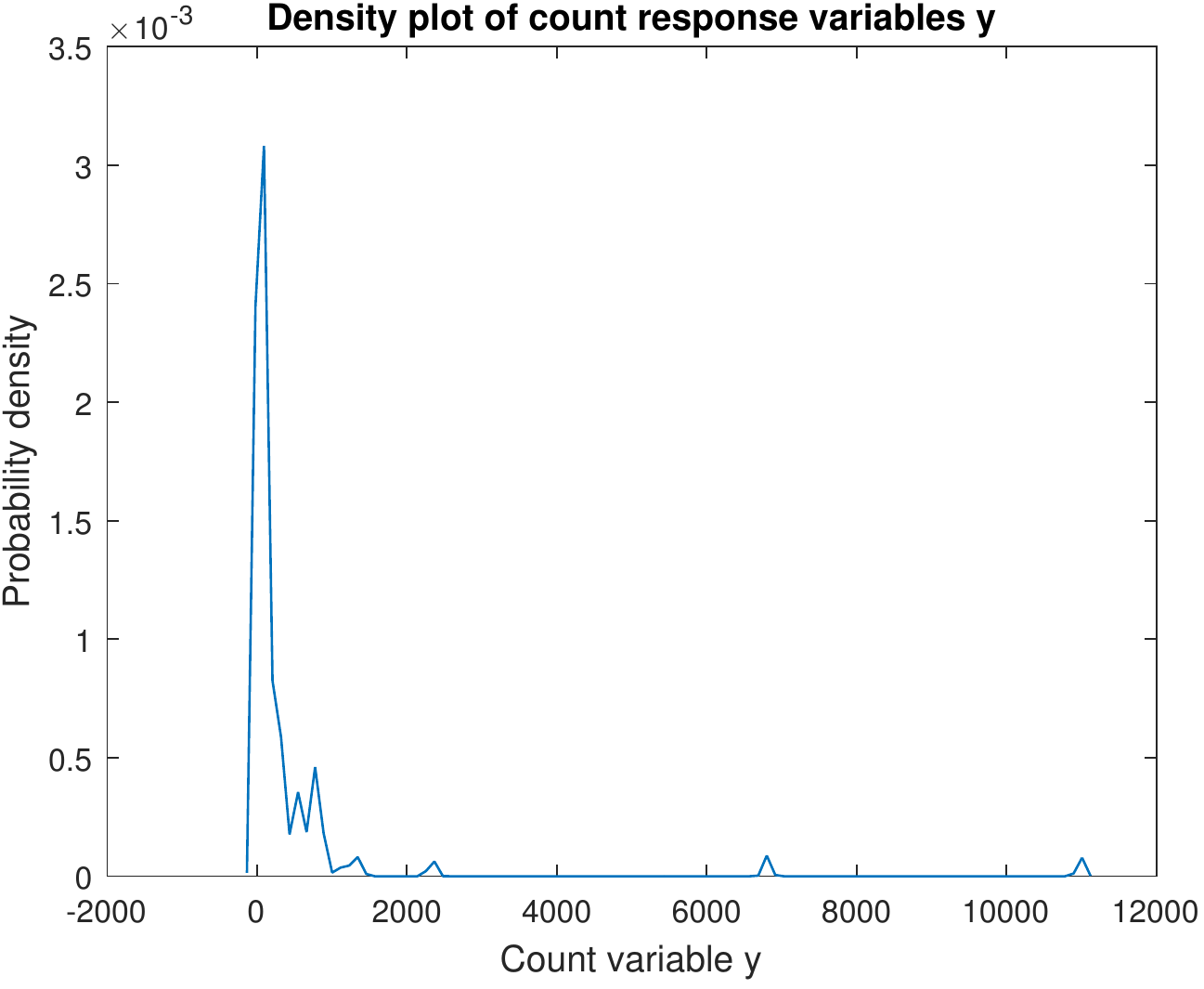}
\caption{Probability density plot of y}
\label{density_responses_sim_2}
\end{subfigure}
\caption{Count response data plots for simulation study 2}
\label{responses_sim_2}
\end{figure}

In this simulation study, we added a random slope term to the model \eqref{sim_1_model}. Therefore, the $\mu$ expression in \eqref{sim_1_mu} is modified as follows:
\begin{equation*} 
\mu_{ij} = \exp \{ x_{1ij}\beta_1 + x_{2ij}\beta_2 + x_{3ij}\beta_3 + \alpha_{1i} + s_{1ij} \alpha_{2i} \}, \hspace{1cm} i = 1,...,20, j = 1,...,5.
\end{equation*}	
we sampled $s_{1ij}$ from unif(0,1) which is the covariate for random slope parameter. Other model specifications and priors are similar to the simulation study 1. The histogram of count response variables and the smoothed kernel density plot is shown in \autoref{responses_sim_2}. The Bayesian quantile regression using Gibbs sampling algorithm proposed above was run for 10000 iterations after initial 2000 burn-in iterations on each jittered data at each quartile. 

\begin{figure}[H]
\includegraphics[width=\textwidth]{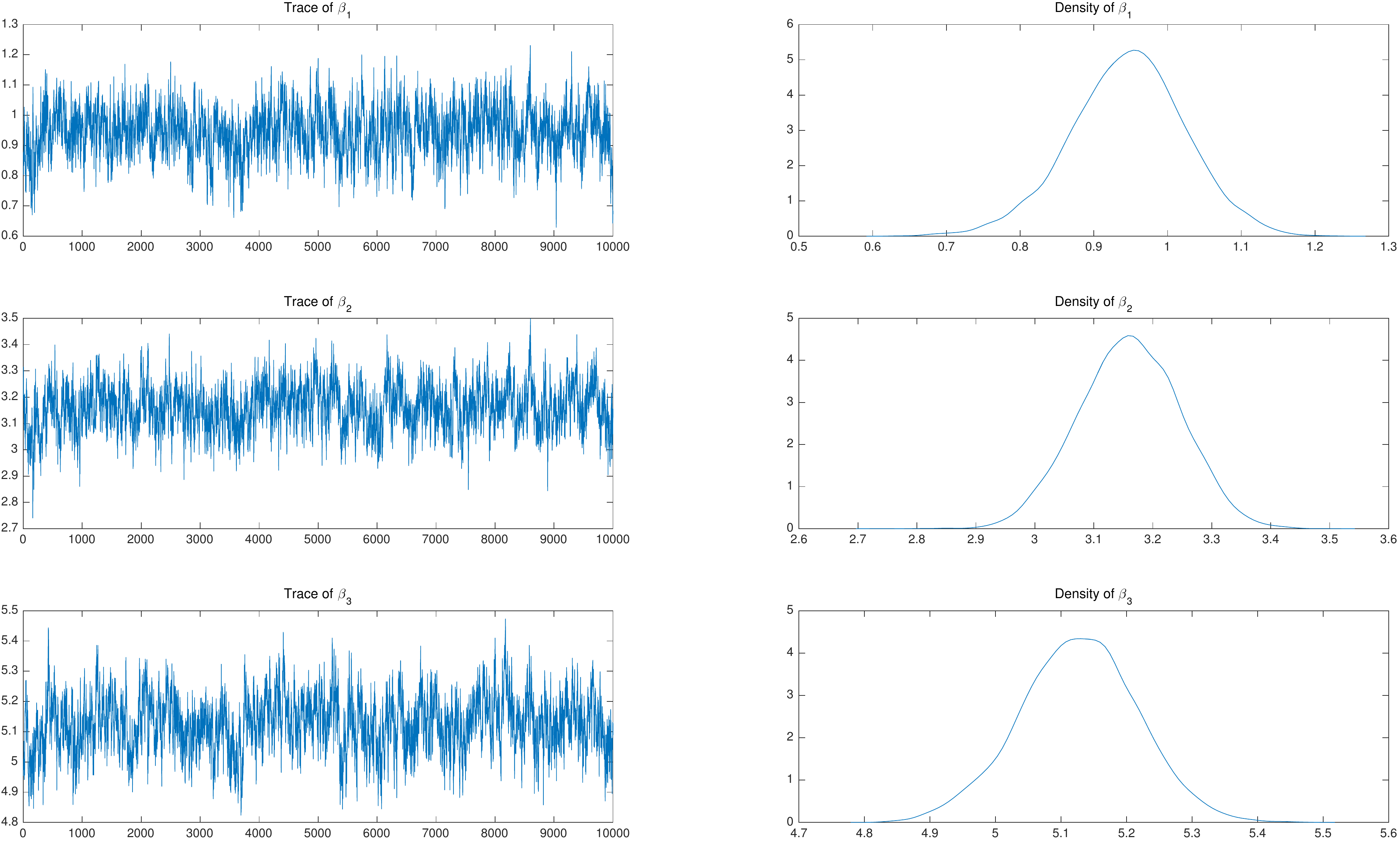}
\centering
\caption{Trace plots and kernel density plots of $\beta_1, \beta_2$ and $\beta_3$ at 0.25 quantile based on 10,000 iterations on one of the jittered dataset in random intercept and slope model}
\label{trace_density_sim_2_quartile_1_beta}
\end{figure}

\begin{figure}[H]
\includegraphics[width=\textwidth]{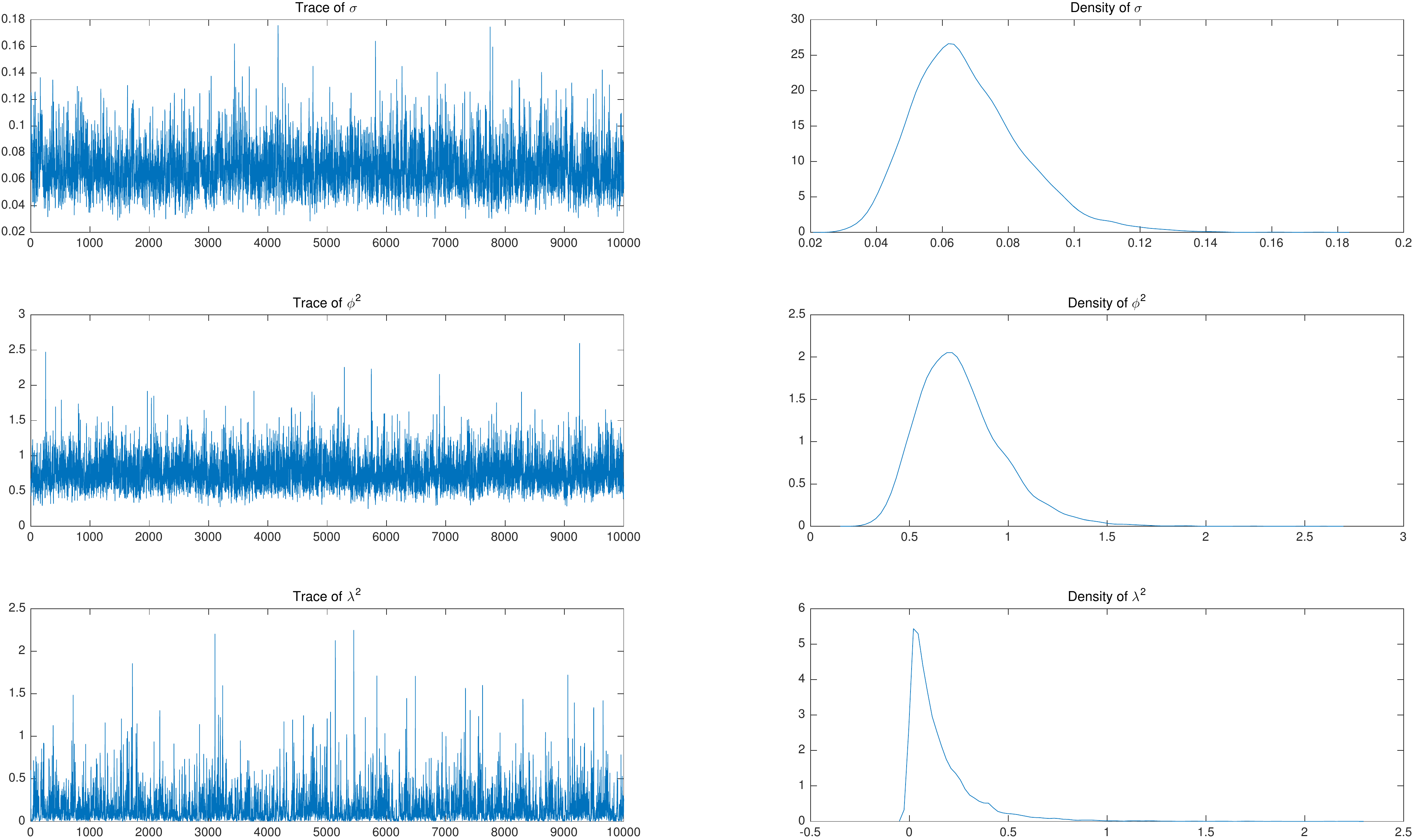}
\centering
\caption{Trace plots and kernel density plots of $\sigma, \phi^2$ and $\lambda^2$ at 0.25 quantile based on 10,000 iterations on one of the jittered dataset in random intercept and slope model}
\label{trace_density_sim_2_quartile_1_hyper}
\end{figure}

The results for each quantile regression model are presented in \autoref{sim_study_2}. Average posterior means for $\beta$ are close to the true values. Standard deviations were calculated according to the formula \eqref{pool_var}. Average credible intervals obtained were smaller in length than any of the individual credible intervals. Trace plots and kernel density plots for $\beta$ and hyperparameters $(\sigma, \phi^2, \lambda^2)$ are given below for $p=0.25$ quantile regression (see \ref{trace_density_sim_2_quartile_1_beta} and 
\ref{trace_density_sim_2_quartile_1_hyper}) 
Graphs at other quartiles can be made available on request. Inspection of trace plots suggests convergence of chains for each parameter and further validates the practical utility of our blocked procedure. 

We also performed the mixed-effects Poisson regression (MEPR) to compare the Bayesian quantile regression models with the standard classical mean regression model. The results for MEPR model are given in \autoref{mean_reg_sim_2}. We can see that coefficient values estimated are close to the true parameter values and they are statistically significant as well. Our model results from \autoref{sim_study_2} are competitive against this frequentist procedure with added benefit of data-driven uncertainty estimates. The NLL for the MEPR model was 464.4 and for the BQRLCD model the values for $1^\text{st}, 2^\text{nd}$ and $3^\text{rd}$ quartiles are 800.5, 515.2 and 606.1 respectively. This shows that the MEPR model performs slightly better than Bayesian quantile regression models. The DIC values for $p=(0.25, 0.50, 0.75)$ quantile models were 2215.4, 1706.6 and 1821.5 respectively which implies that median model provides the best fit among these three quantile regression models.

\begin{table}[H]
\centering
\caption{Average posterior means, standard deviations, and 95\% average credible intervals at $p=0.25, 0.50, 0.75$ quantile levels}
\label{sim_study_2}
\begin{tabular}{@{}lcccccc@{}}
\toprule
Quantile & Par & True $\beta$ & Avg Post Mean & SD & Avg $2.5\%$ & Avg $97.5\%$ \\
\midrule
$p=0.25$ 
& $\beta_1$ & 1 & 0.9481 & 0.0848 & 0.7772 & 1.1088 \\ 
& $\beta_2$ & 3 & 3.1658 & 0.0903 & 2.9881 & 3.3396
\\
& $\beta_3$ & 5 & 5.1310 & 0.0978 & 4.9349 & 5.3165
\\ \midrule
$p=0.50$ 
& $\beta_1$ & 1 & 0.9577 & 0.0656 & 0.8259 & 1.0850 \\ 
& $\beta_2$ & 3 & 3.1089 & 0.0773 & 2.9597 & 3.2612 \\
& $\beta_3$ & 5 & 5.0813 & 0.0847 & 4.9154 & 5.2474 \\ \midrule
$p=0.75$ 
& $\beta_1$ & 1 & 0.9476 & 0.0582 & 0.8319 & 1.0618 \\ 
& $\beta_2$ & 3 & 3.0289 & 0.0717 & 2.8970 & 3.1748 \\
& $\beta_3$ & 5 & 5.0160 & 0.0824 & 4.8572 & 5.1785 \\ 
\bottomrule
\end{tabular}
\end{table}

\begin{table}[H]
\centering
\caption{Mixed Effects Poisson Regression (MEPR) for simulation study 2}
\label{mean_reg_sim_2}
\begin{tabular}{cccccc}
\toprule
Par & True $\beta$ & Coef  & Std Err & \multicolumn{2}{c}{95\% Conf Int} \\ \hline
$\beta_1$ & 1  & 1.0172 & 0.0305  & 0.9574  & 1.0771  \\
$\beta_2$ & 2  & 2.9969 & 0.0454  & 2.9080  & 3.0858  \\
$\beta_3$ & 3  & 5.0304 & 0.0469  & 4.9385  & 5.1224  \\ 
\bottomrule
\end{tabular}
\end{table}


\subsection{Progabide Clinical Trial Data Analysis}

This section presents the application of Bayesian quantile regression for panel count data using the Progabide clinical trial data of 59 epileptic patients. Analyses on this data are reported in \cite{Leppik-et-al-1985} and later in \cite{Thall-Vail-1990}.

\begin{figure}[h]
\begin{subfigure}[h]{0.5\textwidth}
\centering
\includegraphics[width=0.8\textwidth]{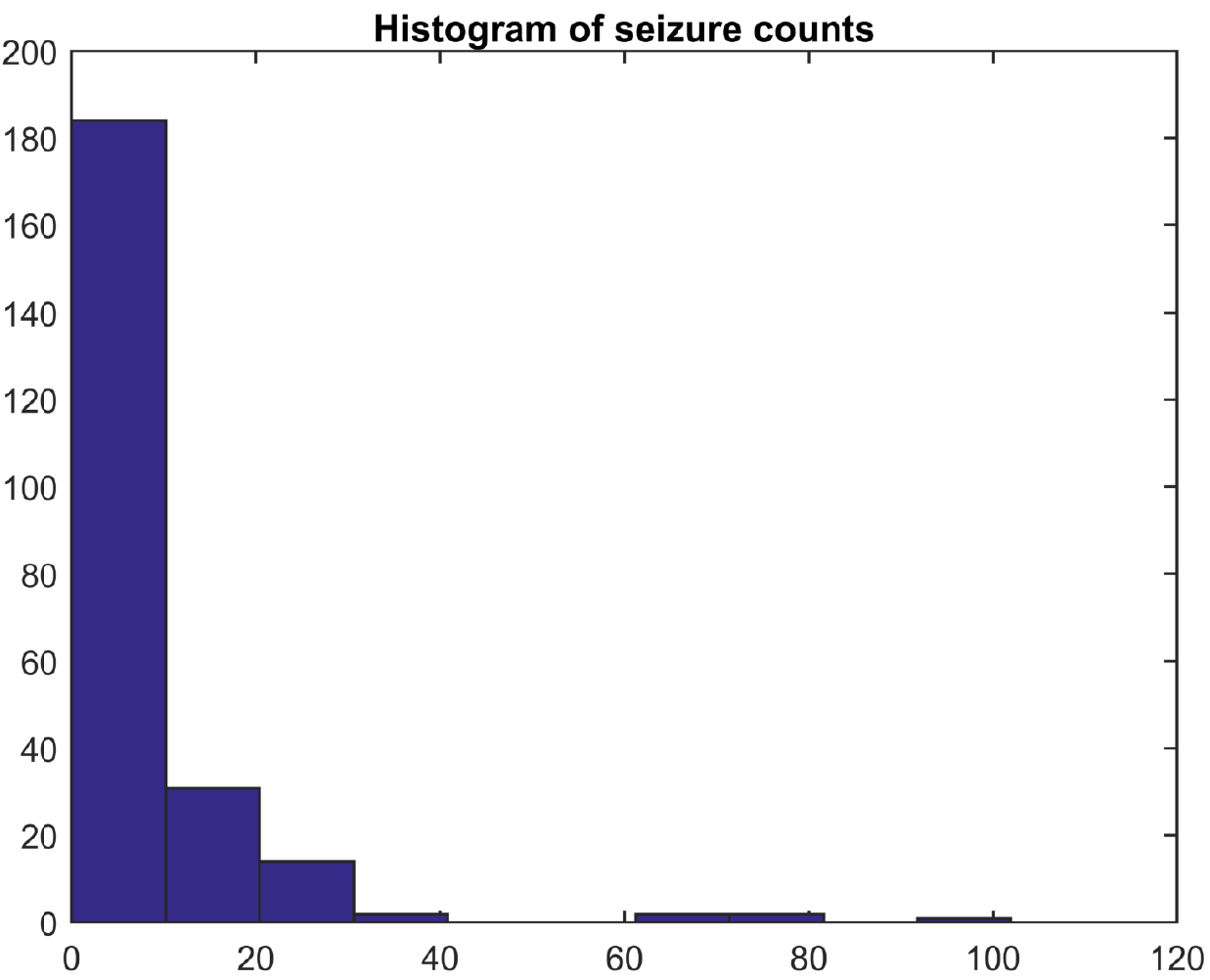}
\caption{Histogram of seizures}
\label{hist_seizure}
\end{subfigure}
\begin{subfigure}[h]{0.5\textwidth}
\centering
\includegraphics[width=0.8\textwidth]{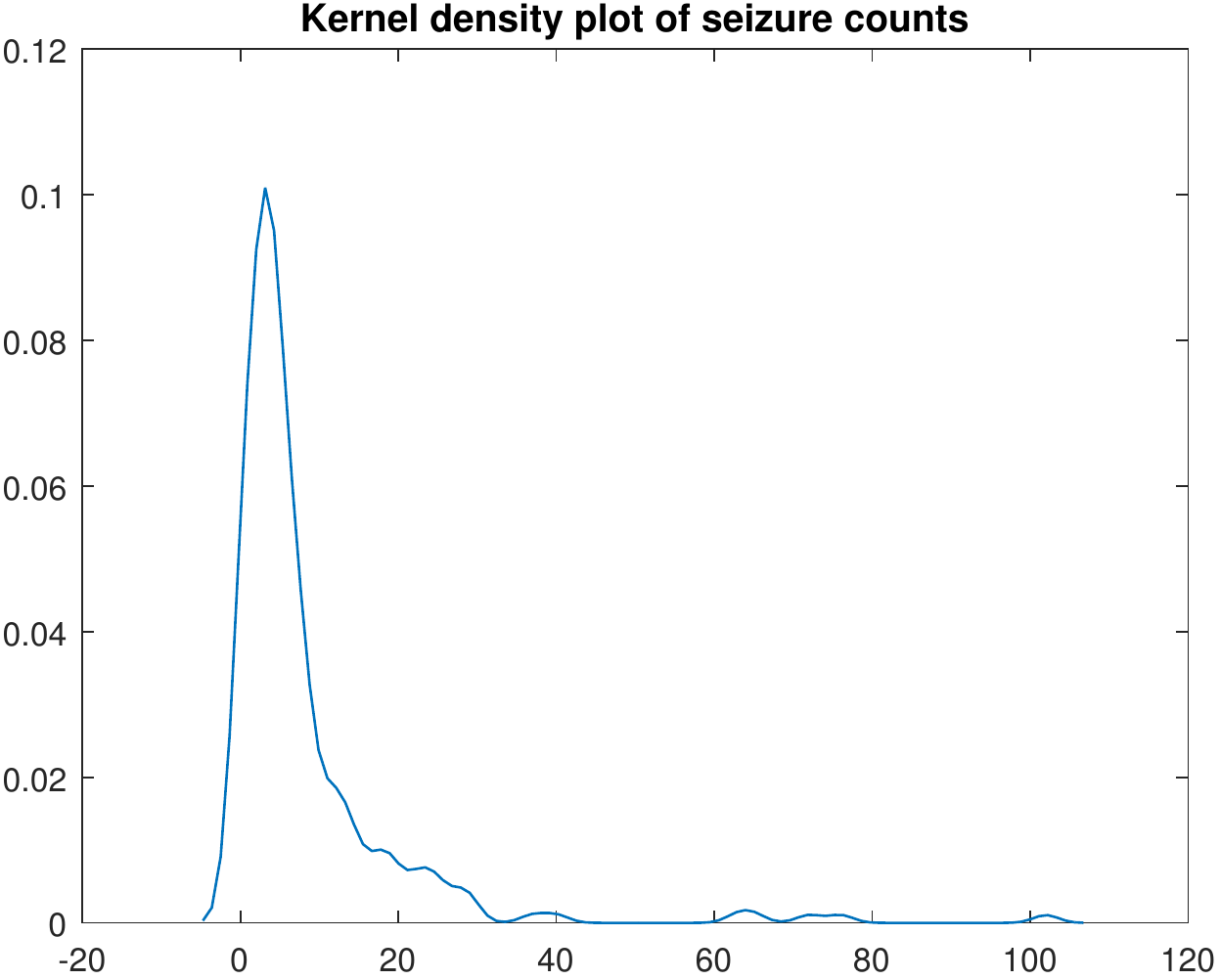}
\caption{Probability density plot of seizures}
\label{density_seizure}
\end{subfigure}
\caption{Biweekly seizure count data plots for Progabide study}
\label{seizure_data}
\end{figure}

Patients with partial seizures were enrolled in a randomized clinical trial of Progabide, an anti-epileptic drug. Participants in the study were randomized to either Progabide/treatment or a placebo/control group, as a chemotherapy adjuvant. Progabide is an anti-epileptic drug and its principal course of action is to improve the gamma-aminobutyric acid (GABA) content. GABA is the primary inhibitory neurotransmitter in the brain. Prior to receiving the treatment, baseline seizure count data on the number of epileptic seizures during the preceding 8-week interval were recorded. At each of four successive post-randomization visits, number of seizures occurring over past 2 weeks was recorded. Although subsequently each patient was crossed over to the other treatment, we have only taken into consideration the pre-crossover responses. The seizure count data exhibits high dispersion, heteroscedasticity, and serial correlation for each patient. Histogram plot and kernel density plot of biweekly seizure counts are given in \autoref{seizure_data}.

We have specified the descriptive statistics of the response variable in \autoref{y_table_1}. We found that the patient with ID = 49 possessed very high baseline as well as the post-randomization seizure counts. Omitting this patient lead to a significant drop in the mean, standard deviation, and correlations of Progabide group. But, this deletion of a patient had no clinical ground and only the prediction purpose was being served. The dispersion as well as the within subject dependence were still at a significant level even after the removal of outlier.    

We considered two longitudinal count data models. The regressors taken in these two models included baseline seizure rate, computed as the natural log of quarter of 8-week pre-randomization seizure count (Base), natural logarithm of age (LnAge), treatment indicator variable for the Progabide group (Trt), last visit indicator variable for the fourth visit to the clinic (Visit) and an interaction term of treatment indicator and baseline seizure count (Base.Trt). The visit indicator variable is added owing to the previous studies showing a significant decrease in the seizures prior to the fourth clinical visit. These variables are a part of larger collection of explanatory variables which were not present in the dataset we retrieved. Selected variables are in-line with the previous literature on the regression analysis of the given data. 

\begin{table}[h]
\centering
\caption{Descriptive statistics of the explanatory variables separated in placebo and Progabide (with and without outlier) groups}
\label{desc_stat}
\begin{tabular}{@{}clccccc@{}}
\toprule
Group & Variables & Mean & Median & Min & Max  & SD \\
\cline{1-7}
\textbf{Placebo} & Age & 28.96 & 29 & 19 & 40  & 5.40  \\
(28 Subjects) & Baseline Seizure Count & 30.79 & 19 & 6  & 111 & 25.63 \\
\midrule
\textbf{Progabide} & Age & 28.71 & 26 & 18 & 57 & 8.79 \\
(31 Subjects) & Baseline Seizure Count & 31.65 & 24 & 7  & 151 & 27.54 \\
\midrule
\textbf{Progabide}& Age & 28.93 & 26 & 18 & 57 & 8.85  \\
(30 Subjects - & Baseline Seizure Count & 27.67 & 23 & 7  & 76 & 17.12 \\
Outlier removed) &&&&&& \\
\bottomrule
\end{tabular}
\end{table}

\begin{table}[h]
\centering
\caption{Descriptive statistics of the biweekly seizure counts post-randomization for placebo and Progabide groups}
\label{y_table_1}
\begin{tabular}{ccccccccccccc}
\hline
      & \multicolumn{6}{c}{\textbf{Placebo} (28 subjects)}        & \multicolumn{6}{c}{\textbf{Progabide} (31 subjects)}       \\ \cline{2-13} 
Visit & Mean  & SD    & \multicolumn{4}{c}{Correlations} & Mean  & SD     & \multicolumn{4}{c}{Correlations} \\ \hline
1     & 9.36  & 9.95  & 1.00   &        &        &       & 8.58  & 17.94  & 1.00    &        &       &       \\
2     & 8.29  & 8.02  & 0.78   & 1.00   &        &       & 8.42  & 11.67  & 0.91    & 1.00   &       &       \\
3     & 8.79  & 14.41 & 0.51   & 0.66   & 1.00   &       & 8.13  & 13.67  & 0.91    & 0.92   & 1.00  &       \\
4     & 8.00  & 7.47  & 0.67   & 0.78   & 0.68   & 1.00  & 6.74  & 11.07  & 0.97    & 0.95   & 0.95  & 1.00  \\ \hline
\end{tabular}
\end{table}

\begin{table}[H]
\centering
\caption{Descriptive statistics of the biweekly seizure counts post-randomization for Progabide group after the outlier is removed}
\label{y_table_2}
\begin{tabular}{ccccccc}
\hline
      & \multicolumn{6}{c}{\textbf{Progabide} (30 Subjects - outlier removed)} \\ \cline{2-7} 
Visit & Mean     & SD       & \multicolumn{4}{c}{Correlations}         \\ \hline
1     & 5.47     & 5.67     & 1.00     &          &          &         \\
2     & 6.53     & 5.51     & 0.45     & 1.00     &          &         \\
3     & 6.00     & 7.25     & 0.63     & 0.70     & 1.00     &         \\
4     & 4.87     & 4.20     & 0.77     & 0.72     & 0.83     & 1.00    \\ \hline
\end{tabular}
\end{table}

We considered random intercept model (model-1) and random intercept with random visit effects model (model-2). We added random effects for the visit indicator as it is a time-varying variable and hence the coefficients on these effects present the subject specific deviations due to the fourth clinic visit. Also, these models were crosschecked for performance with their standard classical counterparts which are the REPR and the MEPR with random visit effects respectively. Priors on the parameters and hyperparameters were same as that specified in the simulation studies. \autoref{bqr_real_data_model} reports the estimates with their standard deviation for 2 models in Bayesian framework obtained from 10000 iterations after initial burn-in of 2000 iterations. On the other hand, the REPR and the MEPR model parameter estimates, their standard errors and p-values are given in \autoref{REPR_real_model} and \autoref{MEPR_real_model} respectively.

The coefficient values of all the regressors except the base and visit variables are statistically insignificant at 5\% significance level for the REPR model. Whereas, in the MEPR model with random visit effects, only the base variable is statistically significant. In the Bayesian setting, besides the base variable, all other variables cannot be differentiated from 0 according to their credible intervals.  We observed that random intercept accounted for significant amount of variability in the REPR and the MEPR model. However, there was less unobserved subject specific variability for the visit indicator variable.

The interaction variable between the treatment and the baseline seizure rate produces interesting results for the classical REPR model. The mean seizure rate for the Progabide group maybe higher or lower than the placebo group depending upon the baseline seizure rate. This indicates contraindication for patients with higher rate of seizures. For other models, due to lack of statistical significance commenting on this phenomenon is irrelevant. 

Finally for model comparison among Bayesian and classical models, NLL values were calculated using the estimates of coefficients. The NLL value for the REPR model was 645.6 and for the model-1 NLL values for $1^\text{st}, 2^\text{nd}$ and $3^\text{rd}$ quartiles were 1044, 789.6 and 692.6 respectively. This indicates that the our method has comparative performance against well established REPR model. Furthermore, the DIC values obtained for these Bayesian models were 2215.5, 1697.8 and 1660.7 respectively which highlights that $p=0.75$ quantile model provides the best fit among these three quantile regression models.

Similar model comparisons were performed for second modeling scenario. The NLL for the MEPR model with random visit effects was 639.8 and for our three Bayesian quartile models the NLL values were 1048.7, 798.4 and 730.4 for $1^\text{st}$, $2^\text{nd}$ and $3^\text{rd}$ quartiles respectively. This again shows that the classical regression model (MEPR in this case) slightly outperforms the best performing Bayesian model. Nonetheless, a frequentist model can be overly confident about its predictions as well as they lack the ability to quantify uncertainty associated with their estimates. Next, the DIC values were 2234.7, 1694.8 and 1702.3 for $1^\text{st}$, $2^\text{nd}$ and $3^\text{rd}$ Bayesian quartile models which suggests that median and $75$the quantile models fit the data similarly well. 

\begin{table}[h]
\centering
\caption{Posterior means and standard deviations of model parameters in  Progabide clinical trial data}
\label{bqr_real_data_model}
\begin{tabular}{llcccccc}
\hline
Model   & Variables & \multicolumn{2}{l}{25th Quartile} & \multicolumn{2}{l}{50th Quartile} & \multicolumn{2}{l}{75th Quartile} \\ \hline
&& Mean & SD & Mean & SD & Mean & SD \\ \cline{1-8} 
\textbf{Model-1} 
& Intercept & -0.1462 & 0.4934 & -0.0634 & 0.3672 & 0.0322 & 0.3693 \\
& Base & 0.8671 & 0.1720 & 0.9100 & 0.1049 & 0.8901 & 0.1025 \\
& Trt & -0.4409 & 0.4153 & -0.2534 & 0.2625 & -0.2259 & 0.2553 \\
& LnAge & -0.0124 & 0.1560 & 0.0698 & 0.1152 & 0.1410 & 0.1167 \\
& Visit & -0.0222 & 0.1883 & -0.0048 & 0.1184 & -0.0512 & 0.1030 \\
& Base.Trt  & 0.0118 & 0.2023 & -0.0561 & 0.1351 & -0.0314 & 0.1323 \\ \cline{1-8} 
\textbf{Model-2} &\textit{(Visit RE)}&&&&&&\\ 
& Intercept & -0.1475 & 0.4990 & -0.0763 & 0.3882 & 0.0069 & 0.3672 \\
& Base & 0.8720 & 0.1710 & 0.9125 & 0.1028 & 0.8891 & 0.0949 \\
& Trt & -0.4275 & 0.4003 & -0.2556 & 0.2543 & -0.2159 & 0.2395 \\
& LnAge & -0.0139 & 0.1546 & 0.0730 & 0.1201 & 0.1477 & 0.1166 \\
& Visit & -0.0421 & 0.1929 & -0.0094 & 0.1220 & -0.0392 & 0.1133 \\
& Base.Trt & 0.0043 & 0.1998 & -0.0565 & 0.1321 & -0.0274 & 0.1255 \\ \hline
\end{tabular}
\end{table}

\begin{table}[h]
\centering
\caption{Random Effects Poisson Regression for Progabide data}
\label{REPR_real_model}
\begin{tabular}{lccc}
\hline
Variables & Coef & Std Err & p-value \\ \hline
Intercept & -0.9362 & 1.0666 & 0.380\\
Base      & 0.8909 & 0.1297  & 0.000\\
Trt       & -0.6410 & 0.4181 & 0.125\\
LnAge     & 0.3602 & 0.3098 & 0.245\\
Visit     & -0.1425 & 0.0590 & 0.016\\
Base.Trt  & 0.1562 & 0.2166 & 0.471\\ \hline
\end{tabular}
\end{table}

\begin{table}[h]
\centering
\caption{Mixed Effects Poisson Regression for Progabide data with random visit effects}
\label{MEPR_real_model}
\begin{tabular}{lccc}
\hline
Variables & Coef & Std Err & p-value\\ \hline
Intercept & -0.9951 & 0.9964 & 0.318 \\
Base      & 0.9032 & 0.1184 & 0.000 \\
Trt       & -0.5964 & 0.3972 & 0.133 \\
LnAge     & 0.3684 & 0.2893 & 0.203 \\
Visit     & -0.0598 & 0.0782 & 0.568 \\
Base.Trt  & 0.1150 & 0.2011 & 0.568 \\ \hline
\end{tabular}
\end{table}


\section{Conclusion} 

\label{Section5} 


In this paper, we present a Bayesian quantile regression model for longitudinal count data and derive an efficient MCMC - Gibbs sampling algorithm to estimate the proposed model. This model makes use of latent variable formulation of count response data and the estimation procedure exploits the normal-exponential mixture representation of asymmetric Laplace density. The developed framework contributes to the literature on quantile regression for discrete data, panel data models for quantile regression, and discrete longitudinal/panel data models. Furthermore it has potential application in the fields like epidemiology, economics and business. One attractive feature our work lies in the robustness of the proposed model which requires minimal assumptions on the distribution of errors allowing for skewed, heavy-tailed distributions for errors as well as accommodation of outliers. The empirical illustrations provided in the paper demonstrate the computational efficiency of the estimation algorithm and the blocking approach.

\clearpage

\clearpage
\appendix


\section{Conditional Densities of the Parameters for Gibbs Sampler} 

\label{AppendixA} 


In the proposed model, the full conditional posterior densities of all the parameters presented in the Gibbs sampling algorithm except the latent data variable $z$ are derived from the complete posterior density expression exhibited in \autoref{joint_latent_density}. From the full hierarchical Bayesian quantile regression model, we get
\begin{align*}
& f(v_{ij}|z_{ij},\beta,\alpha_i,\sigma) \\
& \propto \enskip f(z_{ij}|v_{ij},\beta,\alpha_i,\sigma) \hspace{1mm} f(v_{ij}|\sigma) \\
& \propto \enskip v_{ij}^{-\frac{1}{2}} \hspace{2mm} \exp \left \{ -\frac{1}{2 \tau^2 \sigma v_{ij}}(z_{ij}-x_{ij}^T\beta - s_{ij}^T\alpha_i - \theta v_{ij})^2 - \frac{v_{ij}}{\sigma} \right \} \\
& \propto \enskip  v_{ij}^{-\frac{1}{2}} \hspace{2mm} \exp \left \{ -\frac{1}{2} \left ( \frac{(z_{ij}-x_{ij}^T\beta - s_{ij}^T\alpha_i)^2+\theta^2v_{ij}^2-2\theta v_{ij}(z_{ij}-x_{ij}^T\beta - s_{ij}^T\alpha_i)}{\tau^2 \sigma v_{ij}} + \frac{2 v_{ij}}{\sigma} \right )\right \} \\
& \propto \enskip v_{ij}^{-\frac{1}{2}} \hspace{2mm} \exp \left \{ -\frac{1}{2} \left [ \frac{(z_{ij}-x_{ij}^T\beta - s_{ij}^T\alpha_i)^2}{\tau^2 \sigma} v_{ij}^{-1} + \left ( \frac{\theta^2}{\tau^2 \sigma} + \frac{2}{\sigma} \right ) v_{ij} \right ] \right \} \\ 
& \propto \enskip v_{ij}^{-\frac{1}{2}} \hspace{2mm} \exp \left \{ -\frac{1}{2} \left ( \rho_1 v_{ij}^{-1} + \rho_2 v_{ij}\right ) \right \}
\end{align*}
where, $\theta$ and $\tau$ are defined in \autoref{theta_tau}. The above density expression for an individual $v_{ij}$ can be identified as the kernel of a generalized inverse Gaussian distribution GIG$(\nu_1,\rho_1,\rho_2)$, where, $\nu_1 = \frac{1}{2}$ and,
\begin{align*}
\rho_1 = \frac{(z_{ij}-x_{ij}^T \beta - s_{ij}^T \alpha_i)^2}{\tau^2 \sigma} \quad \text{  and } \quad \rho_2 = \frac{\theta^2}{\tau^2 \sigma} + \frac{2}{\sigma}.
\end{align*}
The full conditional posterior distribution of $\sigma$ denoted by $f(\sigma|z,v,\beta,\alpha,\phi^2)$ is
\begin{align*}
& f(\sigma|z,v,\beta,\alpha)\\
& \propto \enskip f(z|v,\beta,\alpha,\sigma) \hspace{1mm} f(\sigma) \\ 
& \propto \enskip \sigma^{-\frac{n_i N}{2}} \exp \left \{ \sum_{i=1}^N \sum_{j=1}^{n_i} - \frac{1}{2 \tau^2 \sigma v_{ij}}(z_{ij}-x_{ij}^T\beta - s_{ij}^T\alpha_i - \theta v_{ij})^2  \right \} \sigma^{-c_1-1} \exp \left \{ - \frac{c_2}{\sigma} \right \} \\
& \propto \enskip \sigma^{-\frac{n_i N}{2}-c_1-1} \exp \left \{ -\frac{1}{\sigma} \left [ \sum_{i=1}^N \sum_{j=1}^{n_i} - \frac{1}{2 \tau^2 \sigma v_{ij}}(z_{ij}-x_{ij}^T\beta - s_{ij}^T\alpha_i - \theta v_{ij})^2 + c_2 \right ] \right \} \\
& \propto \enskip \sigma^{-\tilde{c}_1-1} \exp \left \{ -\frac{\tilde{c}_2}{\sigma}  \right \}
\end{align*} 
That is the kernel of an inverse gamma distribution with the $\tilde{c}_1$ and $\tilde{c}_2$ parameters given by
\begin{align*}
 \tilde{c}_1 = \frac{n_i N}{2} + c_1 \enskip \text{ and } \enskip \tilde{c}_2 = \frac{1}{2 \tau^2} \sum_{i=1}^N \sum_{j=1}^{n_i} \frac{(z_{ij}-x_{ij}^T\beta - s_{ij}^T\alpha_i - \theta v_{ij})^2}{v_{ij}} + c_2.
\end{align*}
The full conditional distribution of $\beta$ is given as follows:
\begin{align*}
& f(\beta|z,v,\alpha,\sigma,g^2) \\
& \propto \enskip f(z|v,\beta,\alpha,\sigma) \hspace{1mm} f(\beta|g^2) \\
& \propto \enskip \exp \left \{ -\frac{1}{2}  \sum_{i=1}^N \left ( (z_{i}-x_{i}^T\beta - s_{i}^T\alpha_i - \theta v_{i})^T (\tau^2 D_{v_i}^2)^{-1} (z_{i}-x_{i}^T\beta - s_{i}^T\alpha_i - \theta v_{i}) \right ) \right \}  \\ 
& \hspace{10mm} \times \exp \left \{ -\frac{1}{2}\beta^T D_{g^2}^{-1} \beta \right \}  \\
& \propto \enskip \exp \left \{ -\frac{1}{2} \left [ \beta^T \left (\sum_{i=1}^N \frac{x_i^T (D_{v_i}^2)^{-1} x_i}{\tau^{2 n_i}} + D_{g^2}^{-1} \right ) \beta - \beta^T \left ( \sum_{i=1}^N \frac{x_i (D_{v_i}^2)^{-1} (z_{i} - s_{i}^T\alpha_i - \theta v_{i})}{\tau^{2 n_i}}\right ) \right. \right. \\
& \hspace{17mm} \left. \left. - \left  ( \sum_{1=1}^N \frac{(z_{i} - s_{i}^T\alpha_i - \theta v_{i})^T (D_{v_i}^2)^{-1} x_i^T }{\tau^{2 n_i}} \right ) \beta \right ] \right\} \\
& \propto \enskip \exp \left \{ -\frac{1}{2} \left [ \beta^T \tilde{B}^{-1} \beta - \beta^T \tilde{B}^{-1} \tilde{\beta} - \tilde{\beta}^T \tilde{B}^{-1} \beta \right ]\right \},
\end{align*}
where in the third line we collect the terms involving $\beta$ parameter and the successive line introduces two new terms, $\tilde{B}$ and $\tilde{\beta}$, which are stated as follows,
\begin{align*}
 \tilde{B}^{-1} \enskip  =  \enskip \left ( \sum_{i=1}^N \frac{x_i^{T} (D_{v_i}^2)^{-1} x_i}{\tau^{2 n_i}} + D_{g^2}^{-1} \right ) \quad \text{ and}\quad  \tilde{\beta} \enskip  =  \enskip \tilde{B} \left ( \sum_{i=1}^N \frac{x_i^{T}(D_{v_i}^2)^{-1}(z_i-s_i^{T} \alpha_i - \theta v_i)}{\tau^{2 n_i}} \right ) 
\end{align*} 
where, $D_{v_i}$ is the $diag\hspace{0.05cm}(\hspace{0.1cm}\sqrt[]{\sigma v_{i1}}\hspace{0.1cm},...,\hspace{0.1cm}\sqrt[]{\sigma v_{i n_i}}\hspace{0.1cm})$ matrix. Adding and subtracting $\tilde{\beta}^T \tilde{B}^{-1} \tilde{\beta}$ inside the square brackets in the last line of the derivation of distribution of $\beta$ helps in completing the square. 
\begin{align*}
f(\beta|z,v,\alpha,\sigma,g^2) & \propto \enskip \exp \left \{ -\frac{1}{2} \left [ \beta^T \tilde{B}^{-1} \beta - \beta^T \tilde{B}^{-1} \tilde{\beta} - \tilde{\beta}^T \tilde{B}^{-1} \beta + \tilde{\beta}^T \tilde{B}^{-1} \tilde{\beta} - \tilde{\beta}^T \tilde{B}^{-1} \tilde{\beta} \right ]\right \} \\
& \propto \enskip \exp \left \{ -\frac{1}{2} \left [ (\beta-\tilde{\beta})^T \tilde{B}^{-1} (\beta-\tilde{\beta})\right ]\right \}
\end{align*}
As $\tilde{\beta}^T \tilde{B}^{-1} \tilde{\beta}$ does not involve $\beta$ and can be omitted as it gets included in the proportionality constant. The result can be recognized as the kernel of a Gaussian distribution and therefore $\beta|z,v,\alpha,\sigma,g^2 \sim$ N$(\tilde{\beta},\tilde{B})$. 

The full conditional distribution of each $g_h^2$, denoted by $f(g_h^2|\lambda^2)$ is given by
\begin{align*}
f(g_h^2|\beta_h) & \propto \enskip f(\beta_h|g_h^2) \hspace{1mm} f(g_h^2) \\
& \propto \enskip (g_h^2)^{-\frac{1}{2}} \hspace{2mm} \exp \left \{ - \frac{\beta_h^2}{2 g_h^2} \right \} \exp \left \{-\frac{\lambda^2}{2} g_h^2 \right \}  \\ 
& \propto \enskip (g_h^2)^{-\frac{1}{2}} \hspace{2mm} \exp \left \{ -\frac{1}{2} \left [ \beta_h^2 {(g_h^2)}^{-1} + \lambda^2 g_h^2 \right ] \right \} \\ 
& \propto \enskip (g_h^2)^{-\frac{1}{2}} \hspace{2mm} \exp \left \{ -\frac{1}{2} \left [ \rho_3 {(g_h^2)}^{-1} + \rho_4 g_h^2 \right ] \right \} 
\end{align*}
Thus, the above density expression of $g_h^2$ can be recognized as the kernel of a generalized inverse Gaussian distribution, GIG$(\nu_2,\rho_3,\rho_4)$, where $\nu_2 = \frac{1}{2}, \rho_3 = \beta_h^2$ and $\rho_4 = \lambda^2$. 

The full conditional distribution of $\lambda^2$ which is denoted by $f(\lambda^2|g^2)$ is
\begin{align*}
f(\lambda^2|g^2) & \propto \enskip f(g^2|\lambda^2) \hspace{1mm} f(\lambda^2) \\
& \propto \enskip (\lambda^2)^k \exp \left \{ \sum_{h=1}^k -\frac{\lambda^2}{2} g_h^2\right \} (\lambda^2)^{a_1-1} \exp \left \{ -a_2 \lambda^2 \right \} \\ 
& \propto \enskip (\lambda^2)^{k+a_1-1} \exp \left \{ -\lambda^2 \left ( \sum_{h=1}^k \frac{g_h^2}{2} + a_2\right )\right \} \\
& \propto \enskip (\lambda^2)^{\tilde{a}_1-1} \exp \left \{ -\tilde{a}_2\lambda^2 \right \}
\end{align*}
It is clear that the above expression is the kernel of a Gamma distribution, where,
\begin{align*}
 \tilde{a}_1 = k + a_1 \enskip \text{ and } \enskip \tilde{a}_2 = \sum_{h=1}^k \frac{g_h^2}{2} + a_2.
\end{align*}
Th full conditional posterior distribution of $\alpha_i$ is given by
\begin{align*}
& f(\alpha_i|z,v,\beta,\sigma,\phi^2) \\
& \propto \enskip f(z|v,\beta,\alpha,\sigma) \hspace{1mm} f(\alpha_i|\phi^2) \\
& \propto \enskip \exp \left \{ -\frac{1}{2} \sum_{i=1}^N \left ( (z_{i}-x_{i}^T\beta - s_{i}^T\alpha_i - \theta v_{i})^T (\tau^2 D_{v_i}^2)^{-1} (z_{i}-x_{i}^T\beta - s_{i}^T\alpha_i - \theta v_{i}) \right ) \right \}  \\
& \hspace{10mm} \times \exp \left \{ -\frac{\alpha_i^T \alpha_i}{2 \phi^2} \right \}  \\
& \propto \enskip \exp \left \{ -\frac{1}{2} \left [ \alpha_i^T \left (\sum_{i=1}^N \frac{s_i^T (D_{v_i}^2)^{-1} s_i}{\tau^{2 n_i}} + \frac{1}{\phi^2} I_l \right ) \alpha - \alpha_i^T \left ( \sum_{i=1}^N \frac{s_i (D_{v_i}^2)^{-1} (z_{i} - x_{i}^T\beta - \theta v_{i})}{\tau^{2 n_i}}\right ) \right. \right. \\
& \hspace{17mm} \left. \left. - \left  ( \sum_{1=1}^N \frac{(z_{i} - x_{i}^T\beta - \theta v_{i})^T (D_{v_i}^2)^{-1} s_i^T }{\tau^{2 n_i}} \right ) \alpha_i \right ] \right\} \\
& \propto \enskip \exp \left \{ -\frac{1}{2} \left [ \alpha_i^T \tilde{A}^{-1} \alpha_i - \alpha_i^T \tilde{A}^{-1} \tilde{a} - \tilde{a}^T \tilde{A}^{-1} \alpha_i \right ]\right \},
\end{align*}
where, in the third line we omit terms which do not include $\alpha_i$ parameter and on the next line we introduce two new terms, $\tilde{A}$ and $\tilde{a}$, which are defined below,
\begin{align*}
 \tilde{A}^{-1} = \left ( \frac{s_i^{T} (D_{v_i}^2)^{-1} s_i}{\tau^{2 n_i}} + \frac{1}{\phi^2} I_N \right ) \enskip \text{ and } \enskip \tilde{a} = \tilde{A} \left ( \frac{s_i^{T} (D_{v_i}^2)^{-1} (z_i-x_i^{T} \beta - \theta v_i)}{\tau^{2 n_i}}\right ).
\end{align*}
where, $D_{v_i}$ is the $diag\hspace{0.05cm}(\hspace{0.1cm}\sqrt[]{\sigma v_{i1}}\hspace{0.1cm},...,\hspace{0.1cm}\sqrt[]{\sigma v_{i n_i}}\hspace{0.1cm})$ matrix. Adding and subtracting $\tilde{a}^T \tilde{A}^{-1} \tilde{a}$ inside the curly braces in the last line of the derivation of distribution of $\alpha_i$ helps in completing the square.
\begin{align*}
f(\alpha_i|z,v,\beta,\sigma,\phi^2) & \propto \enskip \exp \left \{ -\frac{1}{2} \left [ \alpha_i^T \tilde{A}^{-1} \alpha_i - \alpha_i^T \tilde{A}^{-1} \tilde{a} - \tilde{a}^T \tilde{A}^{-1} \alpha_i + \tilde{a}^T \tilde{A}^{-1} \tilde{a} - \tilde{a}^T \tilde{A}^{-1} \tilde{a} \right ]\right \} \\ 
& \propto \enskip \exp \left \{ -\frac{1}{2} \left [ (\alpha_i-\tilde{a})^T \tilde{A}^{-1} (\alpha_i-\tilde{a})\right ]\right \}
\end{align*}
As $\tilde{a}^T \tilde{A}^{-1} \tilde{a}$ does not involve $\alpha_i$ and can be omitted as it gets included in the proportionality constant. The result can be recognized as the kernel of a Gaussian distribution and therefore $\alpha_i|z,v,\beta,\sigma,\phi^2 \sim$ N$(\tilde{a},\tilde{A})$. 

The full conditional posterior distribution of $\phi^2$ denoted by $f(\phi^2|\alpha)$ is
\begin{align*}
f(\phi^2|\alpha) & \propto \enskip f(\alpha|\phi^2) \hspace{1mm} f(\phi^2) \\ 
& \propto \enskip (\phi^2)^{-\frac{N}{2}} \exp \left \{ -\sum_{i=1}^N  \frac{\alpha_i^T \alpha_i}{2 \phi^2}  \right \} (\phi^2)^{-b_1-1} \exp \left \{ - \frac{b_2}{\phi^2} \right \}  \\
& \propto \enskip (\phi^2)^{-\frac{N}{2}-b_1-1} \exp \left \{ -\frac{1}{\phi^2} \left [ \sum_{i=1}^N  \frac{\alpha_i^T \alpha_i}{2} + b_2 \right ] \right \} \\
& \propto \enskip (\phi^2)^{-\tilde{b}_1-1} \exp \left \{ -\frac{\tilde{b}_2}{\phi^2} \right \}
\end{align*} 
That is the kernel of an inverse gamma distribution with the $\tilde{b}_1$ (shape) and $\tilde{b}_2$ (scale) parameters given as follows:
\begin{align*}
 \tilde{b}_1 = \frac{n_i N}{2} + b_1 \enskip \text{ and } \enskip \tilde{b}_2 = \sum_{i=1}^{n_i}\frac{\alpha_i^T\alpha_i}{2} + b_2.
\end{align*}

\end{document}